# Answering Queries using Views over Probabilistic XML: Complexity and Tractability

Bogdan Cautis
Institut Mines-Télécom, Télécom ParisTech, CNRS LTCI, Paris, France
cautis@telecom-paristech.fr

Evgeny Kharlamov
KRDB Research Centre, Free University of Bozen-Bolzano, Italy
kharlamov@inf.unibz.it

## ABSTRACT

We study the complexity of query answering using views in a probabilistic XML setting, identifying large classes of XPath queries – with child and descendant navigation and predicates – for which there are efficient (PTime) algorithms. We consider this problem under the two possible semantics for XML query results: with persistent node identifiers and in their absence. Accordingly, we consider rewritings that can exploit a single view, by means of compensation, and rewritings that can use multiple views, by means of intersection. Since in a probabilistic setting queries return answers with probabilities, the problem of rewriting goes beyond the classic one of retrieving XML answers from views. For both semantics of XML queries, we show that, even when XML answers can be retrieved from views, their probabilities may not be computable.

For rewritings that use only compensation, we describe a PTime decision procedure, based on easily verifiable criteria that distinguish between the feasible cases – when probabilistic XML results are computable – and the unfeasible ones. For rewritings that can use multiple views, with compensation and intersection, we identify the most permissive conditions that make probabilistic rewriting feasible, and we describe an algorithm that is sound in general, and becomes complete under fairly permissive restrictions, running in PTime modulo worst-case exponential time equivalence tests. This is the best we can hope for since intersection makes query equivalence intractable already over deterministic data. Our algorithm runs in PTime whenever deterministic rewritings can be found in PTime.

## 1. INTRODUCTION

Uncertainty is ubiquitous in data and many applications must cope with it [21]: information extraction from the World Wide Web [13], automatic schema matching in data integration [31], or data-collecting sensor networks [30] are inherently imprecise. This uncertainty is sometimes represented as the *probability* that the data is correct, as with conditional random fields [24] in information extraction, or uncertain schema mappings in [19]. In other cases, only *confidence* in the information is provided by the system, and can be seen after renormalization as approximation of probabilities. It is thus natural to manipulate such probabilistic information in a *probabilistic database management system* [15].

Recent work has proposed models for probabilistic data, both in the relational [35, 16, 23] and XML [28, 22, 2] settings. We focus here on the latter case, which is particularly adapted for the Web. A number of studies on probabilistic XML have dealt with query answering for a variety of models and query languages [1, 28, 2, 22]. At the same time, query optimization over probabilistic data has received little attention. In particular, the problem of answering queries using views, a key approach for optimization, has received no attention so far in both the relational and the semistructured settings. Yet probabilistic query evaluation could greatly benefit from such techniques, as it is often the case that computing probabilistic results is harder than in the deterministic setting.

Views over XML documents can be seen as fragments of data that may be available for further querying. Over a probabilistic document, these data fragments come together with their probability. Given a document $d$, a set of views $v_1, \ldots, v_n$, and a query $q$, the goal is to understand whether one can obtain $q(d)$, the answers of $q$ over $d$, by accessing view results $v_1(d), \ldots, v_n(d)$ only.

For deterministic data, prior research [36, 25] on XPath rewriting studied the problem of equivalently rewriting an XPath query by navigating inside a single materialized XPath view. This would be the only kind of rewriting supported when the query cache can store or obtain only copies of the XML elements in a query answer, while the original node identities are lost. Following a recent industrial trend (supported by systems such as [6]) towards enhancing XPath queries with the ability to expose node identifiers and exploit them via identity-based equality, techniques for multiple-view rewritings built by intersecting several materialized view results were proposed. These are potentially more beneficial, as many queries with no single-view rewriting can be rewritten using multiple views. [8] studied the complexity of rewriting XPath using an intersection of views and described algorithms that apply for any documents and type of identifiers, including application level Ids.

We study in this paper the complexity of answering queries using views in a probabilistic XML setting, identifying large classes of XPath queries for which there are efficient (PTime) algorithms. Polynomial time techniques for view-based rewriting are in our view even more important here than in the deterministic case, given that query evaluation over probabilistic XML is intractable (in combined data and query complexity) [22]. To the best of our knowledge, our work is the first to address this view-based rewriting problem. Since in a probabilistic setting queries return answers with probabilities, the problem of rewriting goes beyond the classical one of retrieving XML answers from views.

*Contributions.* We study the rewriting problem under the two possible semantics for probabilistic XML results: with persistent node Ids and in their absence. Accordingly, we consider alternative plans (rewritings) that can exploit a single view, by means of *compensation*, or plans that can use multiple views, with *intersection*.

We first show that, in the probabilistic setting, the problem of answering queries using views becomes more complex and it does not reduce to its deterministic version. The reason is that query results now involve not only data trees, but also their probabilities. Hence probabilities should also be retrieved from probabilistic view results, by means of a *probabilistic function* computing them.

Even for the simpler setting (without persistent Ids), the existence of the probabilistic function is not guaranteed by the existence of a data-retrieving rewriting. We first present examples of views and queries for which such a function does not exist. Based on a certain notion of probabilistic independence between queries that we introduce (called condition-independence; in short c-independence), we identify the tightest class of queries and views for which this function exists, and we describe how it can be computed efficiently. Before describing the general solution, we discuss in Section 4.3 a particular case that allows (i) a concise and intuitive formulation of the probabilistic function, (ii) an efficient evaluation over the view document, with no (or little) post-processing.

For rewritings with intersection, we first provide a sufficient condition – also based on c-independence – that guarantees that the probabilities of query answers can be computed as a product-like formula over the probabilities of the views appearing in the intersection. For this sound approach, we also present an NP-hardness result for deciding whether a selection of c-independent views for a rewriting is possible. Then, going beyond rewritings that assume c-independent views, we present a sound algorithm, complete under fairly permissive restrictions, whose complexity drops to PTime under widely-applicable assumptions. More precisely, it runs in PTime modulo worst-case exponential time equivalence tests, with this upper-bound being strict whenever deterministic rewritings can be found in PTime. This is the best we can hope for as intersection makes query equivalence intractable already over deterministic data.

All our results are practically interesting as they allow expressive queries and views, with descendant navigation and path filter predicates. For both semantics, the evaluation of an alternative plan is no more expensive than query evaluation over probabilistic XML.

*Outline of the paper.* Preliminaries are given in Section 2. We formalize the view-based rewriting problem for probabilistic XML in Section 3. We then present our results for the two semantics of XML query results: in the absence of persistent node Ids, in Section 4, and in their presence, in Section 5. We discuss other related work in Section 6. We consider possible directions for future work and we conclude in Section 7.

## 2. PRELIMINARIES

We first describe the data and query model, which largely relies on the terminology and notation of [10, 9] and [2]. Minor extensions for probabilistic view-based query rewriting are given in Section 3.1, with the problem statement. For a more detailed presentation of probabilistic XML we refer the reader to [2].

*XML documents.* We assume the existence of a set of labels $\mathcal{L}$ that subsumes both XML tags and values. We consider an XML document as an unranked, unordered rooted tree $d$ modeled by a set of edges $\texttt{edges}(d)$, a set of nodes $\texttt{nodes}(d)$, a distinguished root node $\texttt{root}(d)$ and a labeling function $\mathsf{lbl}$, assigning to each node a label from $\mathcal{L}$. The label of $\texttt{root}(d)$ is called the *document name* of $d$. We assume that each node $n \in \texttt{nodes}(d)$ has a unique *identifier* (e.g., a numeric value) denoted $Id(n)$.

EXAMPLE 1. *Consider the document $d_{\text{PER}}$ in Figure 1 (where* PER *stands for personnel), describing the personnel of an IT department and the bonuses distributed for different projects. The*

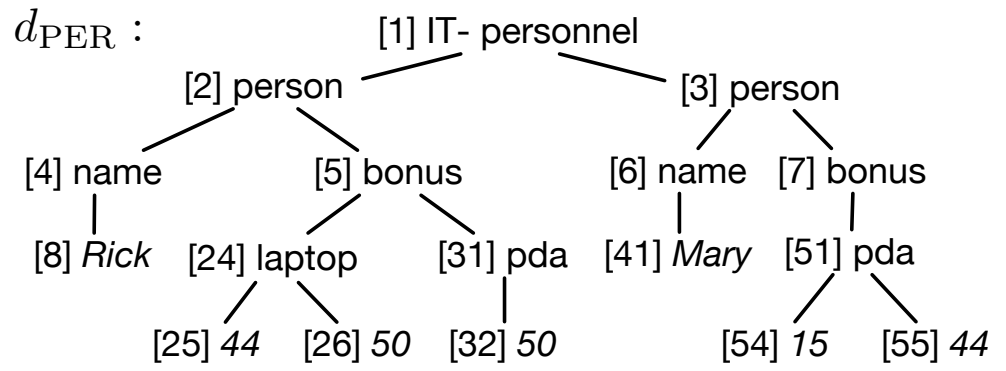

**Figure 1: Example document $d_{\text{PER}}$**

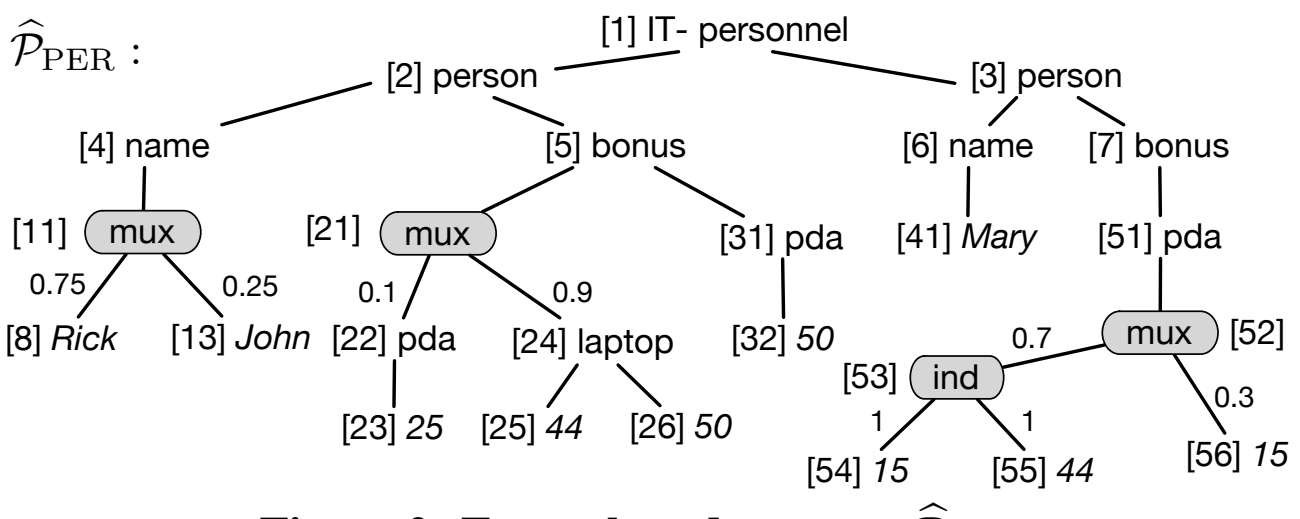

**Figure 2: Example p-document $\widehat{\mathcal{P}}_{\text{PER}}$**

*document $d_{\text{PER}}$ indicates that* Rick *worked under two projects (*laptop *and* pda*) and got bonuses of* 44 *and* 50 *in the former project and* 25 *in the latter one. Identifiers are written inside square brackets and labels are next to them, e.g., the node $n_4$ is labeled* name*, i.e.,* $\mathsf{lbl}(n_4) = \text{name}$.

*Probabilistic documents.* A *finite probability space of XML documents*, or *px-space* for short, is a pair $(\mathcal{D}, \Pr)$ with $\mathcal{D}$ being a set of documents and $\Pr$ mapping every $d \in \mathcal{D}$ to a probability $\Pr(d)$ s.t. $\sum\{\Pr(d) \mid d \in \mathcal{D}\} = 1$.

*p-Documents* [2] give a general syntax for compactly representing px-spaces. Like a document, a p-document is a tree but with two kinds of nodes: *ordinary* nodes, which have labels and are as in documents, and *distributional*, which are used to define the probabilistic process for generating random documents. We consider two kinds of distributional nodes: *mux* (for *mutually exclusive*) and *ind* (for *independent*).

DEFINITION 1. *A* p-document $\widehat{\mathcal{P}}$ *is an unranked, unordered tree with a set of edges $\texttt{edges}(\widehat{\mathcal{P}})$, nodes $\texttt{nodes}(\widehat{\mathcal{P}})$, the root node $\texttt{root}(\widehat{\mathcal{P}})$, and a labeling function $\mathsf{lbl}$, assigning to each node $n$ a label from $\mathcal{L} \cup \{ind(\Pr_n), mux(\Pr_n)\}$. If $\mathsf{lbl}(n)$ is $mux(\Pr_n)$ or $ind(\Pr_n)$, then $\Pr_n$ assigns to each child $n'$ of $n$ a probability $\Pr_n(n')$, and if $\mathsf{lbl}(n) = mux(\Pr_n)$, then also $\sum_{n'} \Pr_n(n') \leq 1$. We require leaves and the root to be $\mathcal{L}$-labeled.*

EXAMPLE 2. *Fig. 2 shows a p-document $\widehat{\mathcal{P}}_{\text{PER}}$ (*PER *stands for personnel) having mux and ind distributional nodes, shown on gray background. Node $n_{52}$ is a mux node with two children $n_{53}$ and $n_{56}$, where $\mathsf{Pr}_{n_{52}}(n_{53}) = 0.7$ and $\mathsf{Pr}_{n_{52}}(n_{56}) = 0.3$.*

A p-document $\widehat{\mathcal{P}}$ has as associated *semantics* a px-space $[\![\widehat{\mathcal{P}}]\!]$ defined by *runs* of the following random process: independently for each $mux(\Pr_n)$ (resp. $ind(\Pr_n)$) node, select at most one (resp. some) of its children $n'$ and delete all other children along with their descendants; then remove in turn each distributional node, connecting ordinary children of deleted distributional nodes with their closest ordinary ancestors. The result of such a run is a *random document* $\mathcal{P}$ (an ordinary document), whose nodes (Ids) are a subset of those of $\widehat{\mathcal{P}}$. Note that there might be several runs resulting in the same $\mathcal{P}$, e.g., by different choices under ordinary nodes of $\widehat{\mathcal{P}}$ that are not kept in $\mathcal{P}$. The *probability of a run* is the product of all *(i)* $\Pr_n(n')$ for each chosen child $n'$ of a *mux* or *ind* node $n$, *(ii)* $1 - \Pr_n(n')$ for each *not* chosen child $n'$ of a *ind* node $n$, *(iii)* $1 - \sum_{n'} \Pr_n(n')$ for all children $n'$ of each *mux* $n$ for which no children were chosen. The probability of a random document $\mathcal{P}$, $\mathsf{Pr}(\mathcal{P})$, is the sum of probabilities of all runs resulting in $\mathcal{P}$.

EXAMPLE 3. *One can obtain the document $d_{\text{PER}}$ from $\widehat{\mathcal{P}}_{\text{PER}}$*

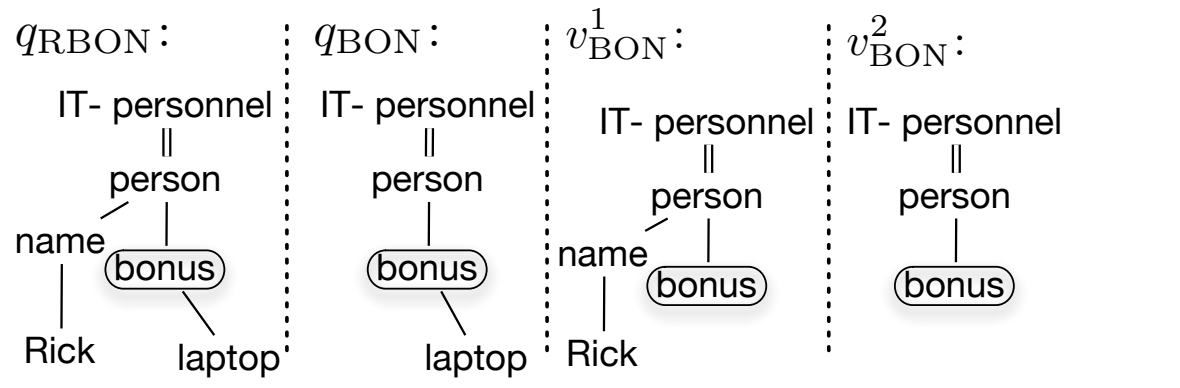

**Figure 3:** TP **queries** $q_{\text{RBON}}$, $q_{\text{BON}}$ **and** TP **views:** $v^1_{\text{BON}}$, $v^2_{\text{BON}}$

*by choosing: the left child of the mux node $n_{11}$, the right child of the mux node $n_{21}$, the left child of the mux node $n_{52}$, and either child of the ind node $n_{53}$. The marginal probability of these choices (and the probability of $d_{\text{PER}}$), is* $0.4725 = 0.75 \times 0.9 \times 0.7 \times 1 \times 1$.

By $d^n$ we denote the subdocument of $d$ rooted at $n$ and by $\widehat{\mathcal{P}}^n$ - the p-subdocument of $\widehat{\mathcal{P}}$ rooted at a node $n$. Note that other kinds of local distributional nodes: *det* (i.e., *deterministic*) and *exp* (i.e., *explicit*) are studied in [2] and all the results of this paper remain valid for p-documents with all four of these distributional nodes. We use only *mux* and *ind* nodes here because they are convenient and the model based on them is a complete representation system [2].

*Tree-pattern queries.* The language of *tree-pattern queries* (TP) is roughly the subset of navigational XPath with child, descendant navigation, predicates, and without wildcards.

Definition 2. *A tree-pattern $q$ is a non-empty, unordered, unranked rooted tree, with a set of nodes $\texttt{nodes}(q)$ labeled with symbols from $\mathcal{L}$, a distinguished node called the* output node $\texttt{out}(q)$ *(i.e., tree-patterns are unary queries), and two types of edges:* child edges*, labeled by / and* descendant edges*, labeled by //. The root of $q$ is denoted $\texttt{root}(q)$.*

The *main branch* $\texttt{mb}(q)$ of $q$ is the path from $\texttt{root}(q)$ to $\texttt{out}(q)$. The *depth* of a main branch node is the distance from it to the root, i.e., the depth of $\texttt{root}(q)$ is 1 and of $\texttt{out}(q)$ is $|\texttt{mb}(q)|$.

For ease of exposition, we often write tree-patterns $q$ in XPath notation [7], and we refer to this notation by $\texttt{xpath(q)}$. We use $\mathsf{lbl}(q)$ as short notation for $\mathsf{lbl}(\texttt{out}(q))$ and the following graphical representation for tree-patterns: we use single lines to denote child edges and double lines for descendant edges, the main branch is the vertical path starting from the root, the output node is in a circle, and predicates are subtrees starting with side branches (see Figure 3). We say that a TP query is formulated *over a document $d$* or *over a p-document $\widehat{\mathcal{P}}$*, if $\mathsf{lbl}(\texttt{root}(q)) = \mathsf{lbl}(\texttt{root}(d))$, respectively $\mathsf{lbl}(\texttt{root}(q)) = \mathsf{lbl}(\texttt{root}(\widehat{\mathcal{P}}))$.

Example 4. *Consider the queries $q_{\text{RBON}}$ and $q_{\text{BON}}$ in Figure 3, left, (where* BON *stands for bonuses and* RBON *for* Rick*'s bonuses). $q_{\text{RBON}}$ asks for bonuses of* Rick *received for the project* Laptop *and $q_{\text{BON}}$ asks for bonuses on* Laptop*. The other two queries $v^1_{\text{BON}}$ and $v^2_{\text{BON}}$ in Figure 3, right, ask for Rick's bonuses and just for bonuses, respectively. The output nodes of all these queries are labeled with* bonus.

The semantics of tree-patterns can be given using embeddings. An *embedding* $e$ of a TP query $q$ into a document $d$ is a function from $\texttt{nodes}(q)$ to $\texttt{nodes}(d)$ satisfying: *(i)* $e(\texttt{root}(q)) = \texttt{root}(d)$; *(ii)* for any $n \in \texttt{nodes}(q)$, $\mathsf{lbl}(e(n)) = \mathsf{lbl}(n)$; *(iii)* for any /-edge $(n_1, n_2)$ in $q$, $(e(n_1), e(n_2))$ is an edge in $d$; *(iv)* for any //-edge $(n_1, n_2)$ in $q$, there is a path from $e(n_1)$ to $e(n_2)$ in $d$.

The *result* of applying a tree-pattern $q$ to a document $d$ is the set:

$$q(d) := \{e(\texttt{out}(q)) \mid e \text{ is an embedding of } q \text{ into } d\}.$$

Example 5. *For the queries in Figure 3, $q_{\text{RBON}}(d_{\text{PER}}) = q_{\text{BON}}(d_{\text{PER}}) = v^1_{\text{BON}}(d_{\text{PER}}) = \{n_5\}$, $v^2_{\text{BON}}(d_{\text{PER}}) = \{n_5, n_7\}$.*

*Intersections of tree-patterns.* We consider in this paper the extension $\text{TP}^\cap$ of TP, denoting *intersections of tree-pattern queries*:

$$\mathsf{TP}^\cap = \{q_1 \cap \cdots \cap q_k \mid k \in \mathbb{N}, q_i \in \mathsf{TP}\}.$$

We say that a $\text{TP}^\cap$ query $q = \bigcap_{i=1}^k q_i$ is formulated *over a set of documents $D$* if $\bigcup_{i=1}^k \mathsf{lbl}(\texttt{root}(q_i)) = \{\mathsf{lbl}(\texttt{root}(d)) \mid d \in D\}$. Its result over $D$ is the node set $\bigcap_{i=1}^k q_i(d \mid d \in D, \mathsf{lbl}(\texttt{root}(q_i)) = \mathsf{lbl}(\texttt{root}(d)))$. Note that *unsatisfiable* $\text{TP}^\cap$ patterns $q$ are possible (when there is no documents $D$ s.t. $q(D) \neq \emptyset$). For the purposes of our paper, we assume hereafter only satisfiable $\text{TP}^\cap$-patterns; satisfiability can be tested in straightforward manner (we refer the reader to this paper's extended version [11], for more details).

*Query equivalence and containment..* A pattern $q_1$ is *contained* in a pattern $q_2$, denoted $q_1 \sqsubseteq q_2$, if $q_1(d) \subseteq q_2(d)$ for every $d$. Also $q_1$ is *equivalent* to $q_2$, $q_1 \equiv q_2$, if $q_1 \sqsubseteq q_2$ and $q_2 \sqsubseteq q_1$. We discuss how to check containment of $\text{TP}^\cap$ queries in Section 5. For TP queries, containment can be decided using containment mappings [4, 27] which are similar to embeddings. In short, a *containment mapping* from $q_1$ to $q_2$ is a function from $\texttt{nodes}(q_1)$ to $\texttt{nodes}(q_2)$ that respects the labels of nodes and maps any two nodes connected with /-edges to nodes connected with /-edges, while nodes connected with //-edges can be mapped to any connected nodes. Then for $q_1$ and $q_2$ in TP, $q_2 \sqsubseteq q_1$ iff there is a containment mapping from $q_1$ to $q_2$ [27]. Note that such a mapping can be computed in polynomial time. For example, observe that $q_{\text{RBON}}$ is contained in $v^2_{\text{BON}}$, and in $q_{\text{BON}}$, $v^1_{\text{BON}}$, while none of the latter two queries is contained in each other.

Unless stated otherwise, in this paper all the TP-queries are assumed to be *minimized*, i.e. without subsumed subqueries that have the same root (minimization can be done in $\mathsf{PTime}$); equivalence of minimized queries amounts to isomorphism [27].

*Querying p-documents.* So far, queries were functions over XML documents, outputting sets of nodes. Over p-documents $\widehat{\mathcal{P}}$, a query $q$ (TP or $\text{TP}^\cap$) naturally yields a set of pairs node-probability $(n, p)$, for $n$ a node of $\widehat{\mathcal{P}}$ and $p$ the probability that $q$ can be embedded into a random document $\mathcal{P}$ of $\widehat{\mathcal{P}}$ by some $e$ s.t. $e(\texttt{out}(q)) = n$; this value ($p$) will also be written as $\Pr(n \in q(\mathcal{P}))$. Formally:

$$q(\widehat{\mathcal{P}}) := \{(n, p) \mid p = \sum\nolimits_{\mathcal{P} \in [\![\widehat{\mathcal{P}}]\!]:\, n \in q(\mathcal{P})} \Pr(\mathcal{P})\}.$$

It is known [22] that TP queries can be evaluated over p-documents $\widehat{\mathcal{P}}$ in $\mathsf{PTime}$ in $|\widehat{\mathcal{P}}|$ (data complexity); the same holds for $\text{TP}^\cap$.

Example 6. *$q_{\text{BON}}$ returns the node $n_5$ iff the right child of the node $n_{21}$ is chosen, thus, $q_{\text{BON}}(\widehat{\mathcal{P}}_{\text{PER}}) = \{(n_5, 0.9)\}$. $v^1_{\text{BON}}$ returns $n_5$ iff the left child of $n_{11}$ is chosen, thus, $v^1_{\text{BON}}(\widehat{\mathcal{P}}_{\text{PER}}) = \{(n_5, 0.75)\}$. $q_{\text{RBON}}$ returns $n_5$ iff both of the above conditions are satisfied, thus, $q_{\text{RBON}}(\widehat{\mathcal{P}}_{\text{PER}}) = \{(n_5, 0.9 \times 0.75)\}$. Since $v^2_{\text{BON}}$ has no predicates, $n_5$ and $n_7$ are labeled with bonuses and are not probabilistically conditioned: $v^2_{\text{BON}}(\widehat{\mathcal{P}}_{\text{PER}}) = \{(n_5, 1), (n_7, 1)\}$.*

## 3. VIEW-BASED REWRITING

We assume a set of *view names* $\mathcal{V}$ disjoint from the set of labels $\mathcal{L}$. By a *view* $v$ we denote a tree-pattern query (that *defines* the view) together with its *name* $v \in \mathcal{V}$.

*Deterministic view-based rewriting.* Let $d$ be a document, $v$ a view. A *(deterministic) view extension* of $v$ over $d$, denoted $d_v$, is an *XML document* obtained by connecting to a root node labeled by a special label *doc*$(v)$ all the documents from the set

$$\{d' \mid d' \textit{ subtree of } d \textit{ s.t. } \texttt{root}(d') \in v(d)\}.$$

Hence $d_v$ can be queried by queries of the form *doc*$(v)/\mathsf{lbl}(v)/\ldots$. If $V$ is a set of views defined over $d$, then $D^d_V = \{d_v \mid v \in V\}$.

For $\mathcal{Q}$ either TP or $\text{TP}^\cap$ and $q \in \mathcal{Q}$ that may use *doc*$(v)/\mathsf{lbl}(v)$, for $v \in V$, the *unfolding of $q$ with $V$*, denoted *unfold*$_V(q)$, is a $\mathcal{Q}$-query obtained from $q$ by replacing each occurrence of $doc(v)/\mathsf{lbl}(v)$ with the definition of $v$.

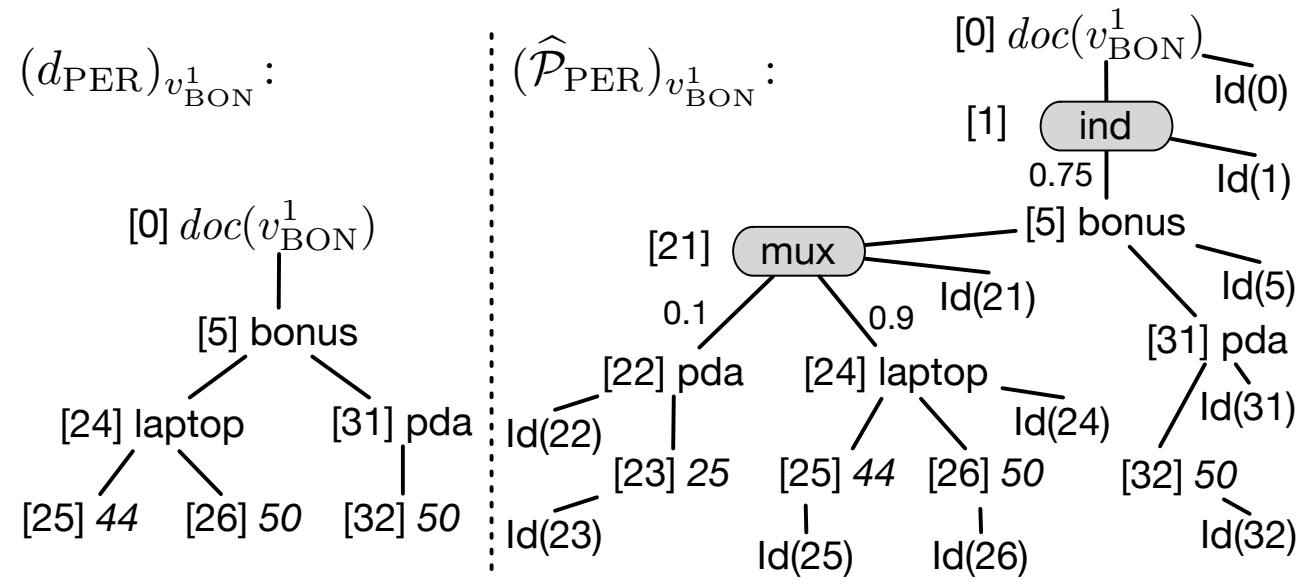

**Figure 4: View extensions $(d_{\mathrm{PER}})_{v^1_{\mathrm{BON}}}$ and $(\widehat{\mathcal{P}}_{\mathrm{PER}})_{v^1_{\mathrm{BON}}}$**

EXAMPLE 7. *$v^1_{\mathrm{BON}}$ and $v^2_{\mathrm{BON}}$ in Figure 3 are views. The view extension $(d_{\mathrm{PER}})_{v^1_{\mathrm{BON}}}$ is in Figure 4, left. Also, $(d_{\mathrm{PER}})_{v^2_{\mathrm{BON}}}$ has a root labeled doc$(v^2_{\mathrm{BON}})$ under which there are the subtrees of $d_{\mathrm{PER}}$ rooted at $n_5$ and $n_7$.*

In the deterministic setting, the problem of query answering using views is to find an alternative query plan $q_r$, called a *rewriting*, that can be used to answer $q$. Formally:

DEFINITION 3. *Let $d$ be a document, $q$ a* TP*-query and $V$ a set of* TP*-views over $d$, $\mathcal{Q} \in \{\mathrm{TP}, \mathrm{TP}^\cap\}$. A deterministic $\mathcal{Q}$-*rewriting *of $q$ using $V$ is a query $q_r \in \mathcal{Q}$ over $D^d_V$ s.t. unfold$_V(q_r) \equiv q$, i.e., for any instance of $d$, unfold$_V(q_r)(d) = q(d)$.*

The two alternatives, TP-*rewritings* and TP$^\cap$-*rewritings*, are respectively motivated by the two possible interpretations of XML query results. In an XML document, nodes have unique Ids used by internal operators (selections, unions, joins, etc.) to manipulate data during query evaluation. Queries can then either *(i)* introduce fresh Ids for the nodes in the result (one for each node Id of the original document), or *(ii)* expose (preserve) in the result the original Ids from the document. The former case corresponds to what is called *the copy semantics*, under which the Ids of any document in $D^d_V$ are disjoint from those of $d$ and from those of any other document in $D^d_V$. Since in this case one cannot know whether nodes in different view extensions are in fact copies of the same node in $d$, the only possible rewritings are those that access a single document from $D^d_V$ and maybe navigate inside it. In the latter case, every document in $D^d_V$ preserves the original Ids, which will identify nodes across different documents in $D^d_V$. One can thus formulate and exploit more complex rewritings, as node Ids can be used to intersect (join by Id) results of different views over the same input data $d$. TP$^\cap$-*rewritings* $q_r$ extend TP-rewritings in that they can access several $D^d_V$ documents at once, by first navigating in individual documents and then intersecting the result.

*View compensation.* As in [10, 9], for TP queries $q_1$ and $q_2$, the result of *compensating* $q_1$ with $q_2$, denoted $\mathtt{comp}(q_1, q_2)$, is a TP-query obtained by deleting the first symbol from $\mathtt{xpath}(q_2)$ and concatenating the rest to $\mathtt{xpath}(q_1)$. $q_2$ is said to be the *compensation* of $q_1$. For example, the result of compensating $q_1 = a/b$ with $q_2 = b[c][d]/e$ is the concatenation of $a/b$ and $[c][d]/e$: $\mathtt{comp}(q_1, q_2) = a/b[c][d]/e$.

Intuitively, compensation brings further navigation over a view's extension and, by results revisited in Section 4 ([36, 3]), a deterministic TP-rewriting will be of form $q_r = \mathtt{comp}(doc(v)/\mathsf{lbl}(v), \dots)$.

## 3.1 Problem Definition

*Encoding probabilistic view extensions.* Let $\widehat{\mathcal{P}}$ be a p-document, $v$ a view. We generalize extensions to the probabilistic case by simply bundling $v$'s results (nothing more, nothing less) in one p-document $\widehat{\mathcal{P}}_v$, which is rooted at a node having a special label $doc(v)$, whose subtree is constructed as follows: (i) plug a unique *ind*-child below $\mathtt{root}(\widehat{\mathcal{P}}_v)$, (ii) for each pair $(\alpha, \beta)$ in the set

$$\{(\widehat{\mathcal{P}}', p) \mid \widehat{\mathcal{P}}' \text{ subtree of } \widehat{\mathcal{P}}, (\mathtt{root}(\widehat{\mathcal{P}}'), p) \in v(\widehat{\mathcal{P}})\},$$

add $\alpha$ as a subtree of this *ind*-node with the probability $\beta$.

The role of $\widehat{\mathcal{P}}_v$ is to give direct access to all the results of the view $v$, by simply evaluating $doc(v)/\mathsf{lbl}(v)$ over this p-document; this does not mean that we assume nor exploit later on an independence property between view outputs (as the *ind*-node may suggest).

A set of p-documents $D^{\widehat{\mathcal{P}}}_V$ for the set of views $V$ and unfolding of a query over $D^{\widehat{\mathcal{P}}}_V$ are defined as in the deterministic case.

Note that, for ease of exposition, we make here a slight abuse of terminology: under both result semantics w.r.t. node Ids, within a view extension, Ids are not necessarily unique; the same Id – either preserved from the original document or a copy of one – may appear several times in the extension. While this is unnecessary in the deterministic context (and could be easily avoided, modulo isomorphic results), it is necessary in the probabilistic one: to compute $\mathsf{Pr}(n \in q(\mathcal{P}))$ for some node $n$, $n$ needs to be properly identified in *all its occurrences* in a view's output, even for TP-rewritings (that use only one view and do not intersect results based on Ids). W.l.g., to simplify the presentation of one of our proofs we make these multiple occurrences directly accessible through queries, by the following post-processing step over view extensions: we plug below each node $n$ a child node with a fresh label "*Id*$(n)$". Also, w.l.g., even under copy semantics, an extension $\widehat{\mathcal{P}}_v$ will be composed of subtrees of the original document instead of copies thereof.

EXAMPLE 8. *Continuing with Example 7, the view extension $(\widehat{\mathcal{P}}_{\mathrm{PER}})_{v^1_{\mathrm{BON}}}$ is in Figure 4, right. Each node $n$ has a new child, labeled "Id$(n)$" (whose own Id is omitted to avoid clutter). Also, $(\widehat{\mathcal{P}}_{\mathrm{PER}})_{v^2_{\mathrm{BON}}}$ has a root labeled doc$(v^2_{\mathrm{BON}})$, with an ind-child under which there are the subtrees of $\widehat{\mathcal{P}}_{\mathrm{PER}}$ rooted at $n_5$ and $n_7$; the edges between this ind-node and its children are valued* 1.

*Probabilistic view-based rewriting.* Query answering using views in the probabilistic setting is more involved than in the deterministic one, as $q(\widehat{\mathcal{P}})$ is a set of node-probability pairs. Therefore, one should deal with two sub-problems: *(i)* find a query in terms of views, that retrieves the nodes $N$ of $q(\widehat{\mathcal{P}})$ (this corresponds to deterministic rewritings) and *(ii)* compute the probabilities for the nodes in $N$, using probabilities from $\mathcal{D}^{\widehat{\mathcal{P}}}_V$. Both sub-problems require algorithms that access p-documents $\mathcal{D}^{\widehat{\mathcal{P}}}_V$ only. Formally:

DEFINITION 4. *Let $\widehat{\mathcal{P}}$ be a p-document, $q$ a* TP *query and $V$ be a set of* TP *views over $\widehat{\mathcal{P}}$, and let $\mathcal{Q} \in \{\mathrm{TP}, \mathrm{TP}^\cap\}$. A* probabilistic *$\mathcal{Q}$-*rewriting *$Q_r = (q_r, f_r)$ of $q$ using $V$ is a pair of*

- *(i) a deterministic $\mathcal{Q}$-rewriting $q_r$ – over random documents $\mathcal{P}$ – of $q$ using $V$, and*
- *(ii) a* probability function *$f_r$ s.t. for every node $n$ of $\widehat{\mathcal{P}}$ it holds that $f_r(n, D^{\widehat{\mathcal{P}}}_V) = \mathsf{Pr}(n \in q(\mathcal{P}))$.*

When $D^{\widehat{\mathcal{P}}}_V$ is clear we will write $f_r(n)$ instead of $f_r(n, D^{\widehat{\mathcal{P}}}_V)$.

Hence the additional challenge that needs to be addressed in probabilistic view-based rewriting is to construct a probability function $f_r$ that, by definition, has access only to the p-documents in $D^{\widehat{\mathcal{P}}}_V$. In Sections 4 and 5 we respectively discuss when and how this is possible for TP and TP$^\cap$-rewritings.

# 4. TP-REWRITINGS

When persistent node Ids cannot be exploited, only one view extension $\widehat{\mathcal{P}}_v$ could be used in a rewriting, by means of navigation [36, 3]. So a deterministic TP-rewriting $q_r$ could only be of the form

$q_r = doc(v)/\mathsf{lbl}(v)[p_1][p_2]..[p_g]$, $q_r = doc(v)/\mathsf{lbl}(v)[p_1][p_2]..[p_g]/p$, *or* $q_r = doc(v)/\mathsf{lbl}(v)[p_1][p_2]..[p_g]//p$.

for $v \in V$ and the TP-queries $p$ and $p_i$ (possibly empty) *compensating* $v$ (with additional predicate conditions and navigation).

*Main ingredients.* Let us fix the names for the main ingredients of this section: the input query $q$ and the view $v$ from the set of views $V$ – to be used in a rewriting – formulated over p-document $\widehat{\mathcal{P}}$, the integer $k = |\mathtt{mb}(v)|$. Let $n$ be one node for which we need to compute the probability $\mathsf{Pr}(n \in q(\mathcal{P}))$ via the $f_r$ function, let $n_1, \ldots, n_a$ be the ancestor-or-self nodes of $n$ that are selected by $v$ (i.e., for which $\mathsf{Pr}(n_i \in v(\mathcal{P})) > 0$).[1]

*Notation for "splitting" queries.* We revisit the terminology of [9], for "splitting" TP queries into a *prefix*, a *suffix*, or several *tokens*. A *prefix* $q'$ of $q$ is any tree-pattern that can be obtained from $q$ by "moving up" the output mark, i.e., by setting as $\mathtt{out}(q')$ a node of $\mathtt{mb}(q)$ and interpreting what follows that node as predicate (side) branches. For any depth $y$, $q^{(y)}$ is the prefix of $q$ with $y$ main branch nodes. A *suffix* $q'$ of $q$ is any subtree of $q$ rooted at a node of $\mathtt{mb}(q)$. The suffix of $q$ rooted at the node of depth $y$ is denoted $q_{(y)}$. The main branch of $q$ can be partitioned by its sub-sequences separated by $//$-edges, and each pattern corresponding to such a sub-sequence is called a *token* of $q$. We can thus see a tree-pattern $q$ as a sequence of tokens $q = t_1// \ldots //t_x$. The token $t_x$, which ends with $\mathtt{out}(q)$, is the *last token* of $q$.

EXAMPLE 9. *The prefix $q^{(2)}_{\mathrm{RBON}}$ corresponds to the XPath $ITpersonnel//person[/name/Rick][bonus/laptop]$, the suffix of $q_{\mathrm{RBON}}$ rooted at the depth-2 node corresponds to $person[name/Rick]/bonus[laptop]$. Also, $q_{\mathrm{RBON}}$ can be split into tokens $t_1 = ITpersonnel$ and $t_2 = person[/name/Rick]/bonus[laptop]$.*

The following three queries that can be obtained from $q$ or $v$ will be often referred to in this section:

- $q'$: the query that can be obtained from the prefix $q^{(k)}$ by removing all predicates of its output node, $\mathtt{out}(q^{(k)})$.
- $v'$: the query that can be obtained from $v$ by removing all predicates of its output node, $\mathtt{out}(v)$.
- $q''$: the query obtained from $q^{(k)}$ by removing all predicates of nodes other than the output node $\mathtt{out}(q^{(k)})$. Formally,

$$q'' = \mathtt{comp}(\mathtt{mb}(q^{(k)}), (q^{(k)})_{(k)}).$$

EXAMPLE 10. *In our running example, for $q_{\mathrm{RBON}}$ as the input query $q$, $v^1_{\mathrm{BON}}$ as the view $v$ ($k = 3$), $q^{(k)}$ is $q$ itself, $q'$ corresponds to the XPath $IT-personnel//person[name/Rick]/bonus$, $q''$ corresponds to $IT-personnel//person/bonus[laptop]$ and $v'$ is $v^1_{\mathrm{BON}}$ itself since there are no predicates on $\mathtt{out}(v)$.*

We start by revisiting a key result from [36, 3], for deterministic rewritings based on compensation:

FACT 1. *[36, 3] Let $q$ and $V$ be* TP*-queries. There exists a deterministic* TP*-rewriting of $q$ over $V$ iff there exists $v \in V$, with $k = |mb(v)|$, such that $\mathtt{comp}(v, q_{(k)}) \equiv q$.*

In our example, we have $\mathtt{comp}(v^1_{\mathrm{BON}}, bonus[laptop]) \equiv q_{\mathrm{RBON}}$.

Fact 1 can be verified in polynomial time [36]. As a reformulation of it, we have that $\mathtt{comp}(v, q_{(k)}) \equiv q$ iff the following hold:

$$q^{(k)} \sqsubseteq v(\sqsubseteq v') \qquad \text{and} \qquad (v \sqsubseteq)v' \sqsubseteq q'.$$

Fact 1 says that, using one view $v$ from $V$, we can find all the nodes $n \in q(d)$ by querying $d_v$ with $q_{(k)}$, i.e., the data in $d_v$ suffices to extract all such $n$. This naturally extends to a probabilistic setting:

[1]Note that only the absence of $//$-edges in the main branch of the view or of the compensation guarantees that $n$'s ancestor-or-self selected by $v$ is unique, regardless of $\widehat{\mathcal{P}}$ (see Definition 5).

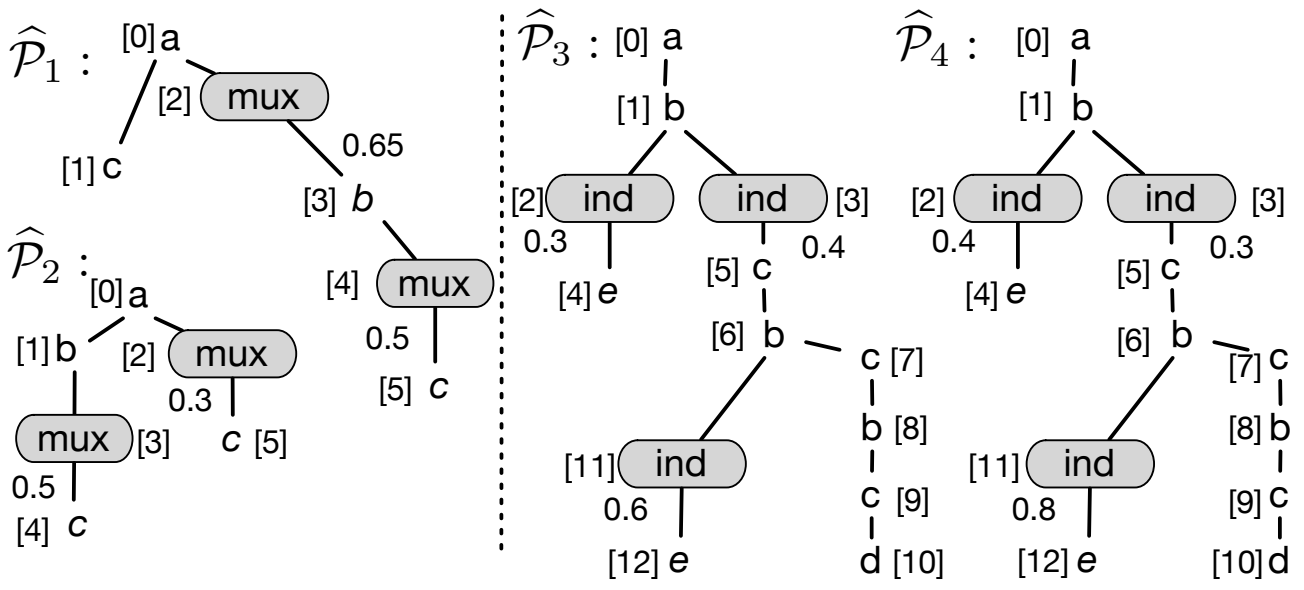

**Figure 5: p-Documents witnessing non-existence of TP-rewritings (Examples 11 and 12)**

we can find all the nodes $n$ from $q(\widehat{\mathcal{P}})$ by querying $\widehat{\mathcal{P}}_v$ with $q_{(k)}$, i.e., the data in $\widehat{\mathcal{P}}_v$ it suffices to extract all $n$s. Note that $n$ is in the query result $q(\mathcal{P})$ iff $\mathsf{Pr}(n \in q(\mathcal{P})) > 0$.

PROPOSITION 1. *Let $q$ and $v$ be* TP*-queries, $k = |mb(v)|$. Let $q_r = \mathtt{comp}(doc(v)/\mathsf{lbl}(v), q_{(k)})$ be a deterministic* TP*-rewriting of $q$ using $v$. Then for every p-document $\widehat{\mathcal{P}}$ the following holds:*

$$\mathsf{Pr}(n \in q(\mathcal{P})) > 0 \quad \textit{if and only if} \quad \mathsf{Pr}(n \in q_r(\mathcal{P}_v)) > 0.$$

The rest of this section is organized as follows. We first discuss the existence of probabilistic TP-rewritings, comparing with the deterministic case and illustrating by examples the aspects on which may depend the construction of the probability-retrieving function $f_r$. These will allow us to articulate the frontier between the feasible cases and the unfeasible ones. We then give some general considerations on which our results for TP-rewritings are built (Section 4.2). Then, before describing the general solution, we discuss in Section 4.3 a particular case - the one of *restricted rewritings* – that allows (i) a concise and intuitive formulation of the $f_r$ component of rewritings, based mainly on probabilistic c-independence, and (ii) an efficient evaluation over the view extension, with no (or little) post-processing. We consider in Section 4.4 the general case, giving one additional necessary condition that along with the ones of Section 4.1 (Proposition 3 therein) will enable a sound and complete procedure for the existence of probabilistic TP-rewritings.

## 4.1 Existence of Probabilistic TP-Rewritings

In the probabilistic setting, we first raise the question: *is information in $\widehat{\mathcal{P}}_v$ always sufficient to extract the probabilities $\mathsf{Pr}(n \in q(\mathcal{P}))$ for nodes $n$ in $q(\widehat{\mathcal{P}})$?* We show that the answer to this question is negative, and that there are $q$ and $v$ for which a deterministic TP-rewriting $q_r$ exists but not a probabilistic one (i.e., no function $f_r$ exists such that for any $\widehat{\mathcal{P}}$ and node $n \in \widehat{\mathcal{P}}$ it holds that $f_r(n) = \mathsf{Pr}(n \in q(\mathcal{P}))$). Thus, the probabilistic rewriting problem is crucially different from the deterministic one. We present two examples (11 and 12) that give insight into this phenomenon.

EXAMPLE 11. *Consider the query $q = a/b[c]$ and the view $v = a[.//c]/b$. We show that there is no probabilistic rewriting $(q_r, f_r)$ for $q$ over $\{v\}$. One can see that $\mathtt{comp}(v, q_{(2)}) = a[.//c]/b[c]$ is equivalent to $q$, so $q_r = \mathtt{comp}(doc(v)/\mathsf{lbl}(v), q_{(2)})$ is a deterministic* TP*-rewriting of $q$ using $v$.*

*Consider now the two p-documents $\widehat{\mathcal{P}}_1$ and $\widehat{\mathcal{P}}_2$ from Figure 5. Clearly, $\mathsf{Pr}(b \in q(\mathcal{P}_1)) = 0.65 \times 0.5$ and $\mathsf{Pr}(b \in q(\mathcal{P}_2)) = 0.5$, and these probabilities are different. The function $f_r$ should compute the first probability $0.325$ on a p-document $(\widehat{\mathcal{P}}_1)_v$ and $0.5$ on $(\widehat{\mathcal{P}}_2)_v$, hence $f_r$ should distinguish these p-documents. However, one can see that these p-documents are* indistinguishable *by $v$: $(\widehat{\mathcal{P}}_1)_v = (\widehat{\mathcal{P}}_2)_v$.[2] Hence, $f_r$ does not exist.*

[2]The probability of the $b$ node is obtained directly as $0.65$ in $(\widehat{\mathcal{P}}_1)_v$, and as $1-(1-0.3)\times(1-0.5) = 0.65$ in $(\widehat{\mathcal{P}}_2)_v$.

The problem exposed in Example 11 comes from the fact that, in the unfolding $a[.//c]/b[c]$, the predicate $[.//c]$ coming from the view (whose probability of matching comes already "packed" into $\widehat{\mathcal{P}}_v$ results, as a condition located above the compensation depth $k$) and the predicate $[c]$ coming from the compensation (whose probability needs to be computed from $\widehat{\mathcal{P}}_v$, as a condition at depth $k$) can *interact*. More precisely, the existence of a match for one predicate depends on the (non-)existence of a match of the other.

***c*-Independent queries.** We now introduce a notion of independence of queries, allowing us to capture necessary properties for the existence of probabilistic rewritings (it will also be used for $\mathrm{TP}^{\cap}$-rewritings). Two TP-queries $q_1$ and $q_2$ are *probabilistically condition-independent* (in short c-independent) – denoted $q_1 \perp q_2$ – if, for every $\widehat{\mathcal{P}}$ and $n \in \widehat{\mathcal{P}}$, $\mathsf{Pr}(n \in (q_1 \cap q_2)(\mathcal{P}))$ equals

$$[\mathsf{Pr}(n \in q_1(\mathcal{P})) \times \mathsf{Pr}(n \in q_2(\mathcal{P}))] \div \mathsf{Pr}(n \in \mathcal{P}).$$

For instance, the queries $q_{\mathrm{BON}}$ and $v^1_{\mathrm{BON}}$ are c-independent, i.e., ($q_{\mathrm{BON}} \perp v^1_{\mathrm{BON}}$). The two TP-queries $a[b]$ and $a[c]$ are not c-independent, as we can easily construct $\widehat{\mathcal{P}}$ over which for some $n \in \mathcal{P}$, we have $\mathsf{Pr}(n \in a[b](\mathcal{P})) > 0$ and $\mathsf{Pr}(n \in a[c](\mathcal{P})) > 0$, yet the probability of the joint test $\mathsf{Pr}(n \in a[b][c](\mathcal{P}))$ equals zero. Deciding probabilistic c-independence is tractable:

PROPOSITION 2. *c-Independence is decidable (*PTime*) in* TP.

PROOF SKETCH. We can use a syntactic notion instead of the probabilistic one for c-independent queries. Intuitively, it precludes dependencies between probabilites of predicates from the two queries to match over p-documents. The two definitions for c-independence can be proven equivalent, and testing for the syntactic one can be done in PTime, in particular via $\mathrm{TP}^{\cap}$-satisfiability tests. Due to space limitations, we refer the reader to the extended version [11] for the complete proof. □

Getting back to Example 11, intuitively, we need to ensure that predicates above depth $k$ from the view $a[.//c]/b$ do not interact with those at depth $k$ from the query. We can capture this in general by testing whether $v' \perp q''$. Note that in the example we have that $v' = a[.//c]/b$ and $q'' = a/b[c]$, and consequently $v' \not\perp q''$. We can prove the following adaptation of Fact 1 to the probabilistic case, towards avoiding unwanted probabilistic interactions.

PROPOSITION 3. *Let $q$ and $V$ be* TP*-queries. Then there exists a probabilistic* TP*-rewriting of $q$ over $V$* only if *there exists $v \in V$ as in Fact 1 – i.e., for $k = |mb(v)|$, $\mathtt{comp}(v, q_{(k)}) \equiv q$ – such that $v' \perp q''$.*

PROOF SKETCH. Example 11 can be used as a generic contradicting construction, to show that when the c-independence condition does not hold there can be no probabilistic rewriting (it illustrates this situation in the simplest possible form). □

An immediate corollary of Proposition 3 is that fewer views may be used to rewrite a query than in the deterministic case:

COROLLARY 1. *[of Prop. 3] $v$ must satisfy $v' \equiv q'$, i.e, $v$ and $q$ are isomorphic modulo predicates at and below the depth $k$.*

We show next that the conditions of Proposition 3 are only necessary for the existence of a probabilistic rewriting in general.

EXAMPLE 12. *Consider the query $q = a//b[e]/c/b/c//d$ and the view $v = a//b[e]/c/b/c$. Clearly, $q_{(5)}$ is a compensation for $v$ and $q_r = \mathtt{comp}(v, q_{(5)})$ is a deterministic* TP*-rewriting for $q$ and $v$. Also, the conditions of Proposition 3 are satisfied by $q$ and $v$.*

*On Figure 5 (right) we present two p-documents $\widehat{\mathcal{P}}_3$ and $\widehat{\mathcal{P}}_4$ that show the non-existence of a probability function $f_r$, such that $(q_r, f_r)$ is a probabilistic* TP*-rewriting for $q$ and $v$.*

*In the two documents, let $n_d$ denoted the node labeled d, let $n_{c1}$ and $n_{c2}$ denote the second node, respectively third node, labeled $c$ (the ones selected by $v$). Clearly, we have that*
$\mathsf{Pr}(n_d \in q(\mathcal{P}_3)) = [0.4 \times 0.3 + 0.6 \times 0.4 - 0.3 \times 0.4 \times 0.6) = 0.288,$
$\mathsf{Pr}(n_d \in q(\mathcal{P}_4)) = [0.3 \times 0.4 + 0.3 \times 0.8 - 0.3 \times 0.4 \times 0.8) = 0.264.$

*A function $f_r$ that would be part of a probabilistic rewriting should be able to compute the probability value for $n_d$ to be selected by $q$ in both $(\widehat{\mathcal{P}}_3)_v$ and $(\widehat{\mathcal{P}}_4)_v$, hence $f_r$ should distinguish these p-documents. But one can see that these p-documents are* indistinguishable *by $v$ (i.e., $(\widehat{\mathcal{P}}_3)_v = (\widehat{\mathcal{P}}_4)_v$) as in both documents $n_{c1}$ is selected by $v$ with probability $0.12$, while $n_{c2}$ is selected by $v$ with probability $0.24$. Hence, $f_r$ does not exist.*

*The reason for which it is not possible to retrieve the right probability values for q's answer is twofold: (i) there are multiple (two) c-ancestors of the d-node, whose probabilities would have to be taken into account for $\mathsf{Pr}(n_d \in q(\mathcal{P}))$, and (ii) the images of the $b[e]/c/b/c$ part of $v$ (i.e., its last token) are not necessarily disjoint. This is because the sequence of labels $(b, c, b, c)$ has a prefix – $(b, c)$ – that is also a suffix thereof; we call it hereafter a* ***prefix-suffix****. As a consequence, in particular, the separate probability of the lower image of $b[e]/c/b/c$ is not computable because the $[e]$ predicate might match in a part of the document that is never visible in view results (we only have access to $n_{c1}$ and its descendants).*

The above example illustrates the remaining aspects on which the existence of a probabilistic rewriting may depend. These have to do with the nodes $n_1, \ldots, n_a \in v(\widehat{\mathcal{P}})$ one may have to inspect to compute $\mathsf{Pr}(n \in q(\mathcal{P}))$ (whether there is just one such ancestor-or-self node or there are several) and the last token of $v$ (whether it has predicates, whether images of it in a document are always disjoint). These aspects will allow us to fully characterize the feasible cases, in Sections 4.3 and 4.4. We first give some general considerations on which our results for TP-rewritings are built.

## 4.2 General Results

To start, we can always formulate $\mathsf{Pr}(n \in q(\mathcal{P}))$ as follows:

$$\mathsf{Pr}(n \in q(\mathcal{P})) = \mathsf{Pr}(\bigvee_{i=1}^{a} [n_i \in v'(\mathcal{P}) \ \wedge \ n \in q_{(k)}(\mathcal{P}^{n_i})]).$$

Recall $a$ is the number of $n$'s ancestor-or-self nodes selected by $v$. So we can always formulate $f_r$ in terms of view extensions as:

$$f_r(n) = \mathsf{Pr}(\bigvee_{i=1}^{a} [n_i \in v'(\mathcal{P}) \ \wedge \ n \in q_{(k)}(\mathcal{P}_v^{n_i})]).$$

Let $e_i$ denotes the event $n_i \in v'(\mathcal{P}) \ \wedge \ n \in q_{(k)}(\mathcal{P}_v^{n_i})$, for $i = 1, a$. By the inclusion-exclusion principle, we can give the following general formulation of the $f_r$ function:

LEMMA 1. *$f_r$ can always be formulated as follows:*

$$f_r(n) := \mathsf{Pr}(\bigvee_{i=1}^{a} e_i) = \sum_i \mathsf{Pr}(e_i) - \sum_{i1,i_2} \mathsf{Pr}(e_{i_1} \cap e_{i_2}) + \ldots + \ldots (-1)^{r-1} \mathsf{Pr}(\bigcap_i e_i). \quad (1)$$

Under the independence condition $v' \perp q''$, the following holds:

LEMMA 2. *Under the conditions of Proposition 3 – i.e., for $k = |mb(v)|$, $\mathtt{comp}(v, q_{(k)}) \equiv q$ and $v' \perp q''$ – we can compute the probability of an event $e_i$ as follows:*

$$\mathsf{Pr}(e_i) = [\mathsf{Pr}(n_i \in v(\mathcal{P})) \div \mathsf{Pr}(n_i \in v_{(k)}(\mathcal{P}_v^{n_i}))] \times \times \ \mathsf{Pr}(n \in q_{(k)}(\mathcal{P}_v^{n_i})). \quad (2)$$

PROOF SKETCH. Immediate by the following reformulations:

$$\mathsf{Pr}(e_i) = \mathsf{Pr}(n_i \in v'(\mathcal{P})) \times \mathsf{Pr}(n \in q_{(k)}(\mathcal{P}_v^{n_i}) \mid n_i \in v'(\mathcal{P})) = \mathsf{Pr}(n_i \in v'(\mathcal{P})) \times \mathsf{Pr}(n \in q_{(k)}(\mathcal{P}_v^{n_i})).$$

$\mathsf{Pr}(n_i \in v'(\mathcal{P})) = [\mathsf{Pr}(n_i \in v(\mathcal{P})) \div \mathsf{Pr}(n_i \in v_{(k)}(\mathcal{P}_v^{n_i}))]$. □

Note that Lemma 1 does not imply that $f_r$ is always computable (recall Example 12). In fact, we will show in Section 4.4 that the probability of *joint $e_i$ events* may not be computable from $\widehat{\mathcal{P}}_v$. Before discussing this, we consider a restricted case (where $v$ selects an unique ancestor-or-self node of $n$, i.e., $a = 1$), for which there is no need to manage joint events.

## 4.3 Restricted TP-Rewritings

DEFINITION 5. *We say that a* TP-*rewriting using a view $v$ and compensation $c$ is* restricted *if either $\mathtt{mb}(v)$ has no //-edges, or $\mathtt{mb}(c)$ has no //-edges.*

The case of restricted rewritings was brought forward by the following question: *In the deterministic case, the XML answers $q(d)$ can be obtained by executing the alternative plan over the view extension $d_v$ (a deterministic XML document); since, by construction, in the probabilistic case, the view extension $\widehat{\mathcal{P}}_v$ is a p-document, would it be enough to query it with the rewriting and get "for free" both the XML nodes $n$ of $q(\widehat{\mathcal{P}})$ and their probabilities $\mathsf{Pr}(n \in q(\mathcal{P}))$?* This would represent, in our view, the most intuitive formulation of the $f_r$ function, requiring no post-processing after the querying phase. We show that this is indeed possible for restricted rewritings under Proposition 3's conditions. More precisely, we show that this approach works modulo one minor adjustment: account for certain probability values (of $\mathtt{out}(v)$ predicates to match), which come already "packed" into results of $v$; these have to be divided away since they will be re-accounted for by compensation.

Let $n_a$ be the unique ancestor-or-self of $n$ that is selected by $v$. The $f_r$ formulation from Eq. (1) can be now simplified, reflecting the following: the probability $\mathsf{Pr}(n \in q(\mathcal{P}))$ that $n$ occurs in $q$'s result amounts to the probability $\mathsf{Pr}(n \in q_r(\mathcal{P}_v))$ that $n$ is selected by $q_r$ in $\mathcal{P}_v$, divided by the probability $\mathsf{Pr}(n_a \in v_{(k)}(\mathcal{P}_v^{n_a}))$ that $n_a$ verifies the predicates on $\mathtt{out}(v)$.

THEOREM 1. *Let $q_r = \mathtt{comp}(doc(v)/\mathsf{lbl}(v), q_{(k)})$ be a restricted deterministic* TP-*rewriting of $q$ using $v$. Then, there exists a probabilistic* TP-*rewriting $(q_r, f_r)$ of $q$ using $v$ if and only if $v' \perp q''$. Moreover, $f_r$ can be computed as follows:*

$$\mathsf{Pr}(n \in q(\mathcal{P})) = \mathsf{Pr}(n \in q_r(\mathcal{P}_v)) \div \mathsf{Pr}(n_a \in v_{(k)}(\mathcal{P}_v^{n_a})).$$

PROOF SKETCH. We start by assuming that $v' \perp q''$, which was already proven a necessary condition. We discuss next how $f_r$ can be computed when this c-independence condition holds.

Let us first assume that there are no predicates on $\mathtt{out}(v)$, which implies that $v = v'$. Hence, for the assumed node $n_a$, we have that $\mathsf{Pr}(n_a \in v_{(k)}(\mathcal{P}^{n_a})) = 1$. Knowing this, let us now understand whether the equality we aim for can indeed hold, namely:

$$\mathsf{Pr}(n \in q(\mathcal{P})) = \mathsf{Pr}(n \in q_r(\mathcal{P}_v)).$$

By $\widehat{\mathcal{P}}_v$'s definition, we can write the right-hand side as follows:

$$\begin{aligned}\mathsf{Pr}(n \in q_r(\mathcal{P}_v)) =& \mathsf{Pr}(n_a \in doc(v)/\mathsf{lbl}(v)(\mathcal{P}_v)) \times \\ & \mathsf{Pr}(\textit{of matching the rest of } q_r \textit{ from } n_a \textit{ in } \mathcal{P}_v) \\ =& \mathsf{Pr}(n_a \in v(\mathcal{P})) \times Pr(n \in q_{(k)}(\mathcal{P}_v^{n_a})) \\ =& \mathsf{Pr}(n_a \in v(\mathcal{P})) \times Pr(n \in q_{(k)}(\mathcal{P}^{n_a})) \\ =& \mathsf{Pr}(n \in q(\mathcal{P})).\end{aligned}$$

The first two reformulations are immediate by $\mathcal{P}_v$'s construction, the third one is enabled by the c-independence condition $v' \perp q''$.

To complete the proof, if there are predicates on $\mathtt{out}(v)$, by the same c-independence condition we have that

$$\mathsf{Pr}(n_a \in v'(\mathcal{P})) = \mathsf{Pr}(n_a \in v(\mathcal{P})) \div \mathsf{Pr}(n_a \in v_{(k)}(\mathcal{P}^{n_a})),$$

and then we can use $v'$ in the previous line of reasoning. □

As an immediate corollary of Theorem 1, when $v$ has no predicates on the output node, the $f_r$ function becomes the intuitive one:

$$\mathsf{Pr}(n \in q(\mathcal{P})) = \mathsf{Pr}(n \in q_r(\mathcal{P}_v)),$$

hence a simple interrogation of the view extension with the plan $\mathtt{comp}(doc(v)/\mathsf{lbl}(v), q_{(k)})$ would retrieve both the XML data and (for free) their probability.[3]

EXAMPLE 13. *A deterministic* TP-*rewriting for $q_{\mathrm{BON}}$ is $q_r = \mathtt{comp}(doc(v^2_{\mathrm{BON}})/bonus, q_{(3)})$, and this plan is obviously restricted. Observe that $(v^2_{\mathrm{BON}})' \perp (q_{\mathrm{BON}})''$ (and $(v^2_{\mathrm{BON}})' \equiv (q_{\mathrm{BON}})'$), so Theorem 1 applies. So the probability $\mathsf{Pr}(n_5 \in q_{\mathrm{BON}}(\mathcal{P}_{\mathrm{PER}}))$ is $\mathsf{Pr}(n_5 \in q_r((\mathcal{P}_{\mathrm{PER}})_{v^2_{\mathrm{BON}}})) \div \mathsf{Pr}(n_5 \in (v^2_{\mathrm{BON}})_{(3)}(\mathcal{P}^{n_5}_{\mathrm{PER}})) = 0.9 \div 1$. Besides $n_5$, for all other nodes $n_i$, $\mathsf{Pr}(n_i \in q_{\mathrm{BON}}(\mathcal{P}_{\mathrm{PER}}))$= 0 since $\mathsf{Pr}(n_i \in q_r((\mathcal{P}_{\mathrm{PER}})_{v^2_{\mathrm{BON}}}))$= 0.*

## 4.4 Unrestricted TP-Rewritings

We now consider the general case, starting with an additional necessary condition that, along with the ones of Proposition 3, enables a sound and complete procedure for the existence of probabilistic rewritings. In order to go beyond the scope of the previous section (i.e., restricted rewritings), we must assume that (i) the view $v$ has at least one //-edge in the main branch, and (ii) the compensation $q_{(k)}$ has at least one //-edge in the main branch.

We show that the remaining ingredient for deciding whether a probabilistic TP-rewriting exists is the *last token* of $v$. Let $t$ be this token, of the form $t = l_1[Q_1]/\ldots/l_m[Q_m]$, where $l_m = \mathsf{lbl}(v)$ and any of $Q_1, \ldots, Q_m$ may be empty. Also, let $u$ denote the length of the maximal *prefix-suffix* of the sequences of labels $(l_1, \ldots, l_m)$, so $0 \leq 2 \times u \leq m$. Hence, when $u \geq 1$, we can write $t$ as follows:

$$l_1[Q_1]/l_2[Q_2]/\ldots/l_u[Q_u]/\ldots/l_1[Q_{m-u+1}]/\ldots/l_{u-1}[Q_{m-1}]/l_u[Q_m]$$

EXAMPLE 14. *Revisiting Example 12, the last token of our view is $b[e]/c/b/c$, for which the sequence of labels of the main branch, $(b, c, b, c)$, has a maximal prefix-suffix of length 2. Hence in the example we have $u = 2$.*

We can now give our main result for unrestricted rewritings.

THEOREM 2. *Let $q$ and $v$ be* TP-*queries s.t. there is a deterministic, non-restricted* TP-*rewriting $q_r = \mathtt{comp}(doc(v)/\mathsf{lbl}(v), q_{(k)})$. Let $u$ be the size of the maximal prefix-suffix of $v$'s last token. There exists a probabilistic* TP-*rewriting $(q_r, f_r)$ of $q$ using $v$ iff*

1. *$v' \perp q''$ (the condition of Proposition 3), and*
2. *the first $u-1$ nodes of $v$'s last token have no predicates.*

*Moreover, $f_r$ can be computed as in Equation (1).*

PROOF SKETCH. Example 12 can be used as a generic contradicting construction, to show that when some of the first $u-1$ nodes of $v$'s last token have predicates there can be no probabilistic rewriting (observe that it illustrates this situation in the simplest possible form). We describe in the rest of the proof one possible way to build $f_r$ when the condition holds, via queries that exploit the special $Id(n)$ nodes we introduced in view extensions.

The individual probability of $e_i$ events can be computed as in Eq. (2). We detail next the probability of *joint $e_i$ events.*

**Case $\mathbf{u = 0}$**. When the label sequence $(l_1, \ldots, l_m)$ has no prefix-suffix, any probability of the form $\mathsf{Pr}(e_i \cap e_j)$, for $n_i$ ancestor of $n_j$, can be computed as follows.

[3] Note that, for any p-document $\widehat{\mathcal{P}}$, any node $n$ s.t. $\mathsf{Pr}(n \in q_r(\mathcal{P}_v)) > 0$, if there is only one ancestor-or-self $n_a$ of $n$ s.t. $\mathsf{Pr}(n_a \in v(\mathcal{P})) > 0$, Theorem 1's approach is sound and complete, regardless of whether $q_r$ is restricted or not. However, in the case of unrestricted plans, this approach would be data-dependent.

Assuming that $n_i \in v'(\mathcal{P})$ already, we construct a TP$^\cap$-pattern $\alpha$ that will test in $\widehat{\mathcal{P}}$ the remaining conditions for $e_i \cap e_j$:

$$\alpha = q_{(k)} \cap \texttt{comp}(l_m//l_1[Q_1]/\dots/l_m[Q_m][Id(n_j)], q_{(k)}).$$

Knowing $n_i \in v'(\mathcal{P})$, we can then test by $n \in \alpha(\mathcal{P}^{n_i})$ all the remaining conditions for $e_i \wedge e_j$. More precisely, we test that:

- $n_j$ is also selected by $v'$; for this, only the conditions of the last token need to be tested since the rest matches already for $n_i$ to be selected by $v'$,
- $n$ is selected by $q_{(k)}$ from $n_i$ (left-hand side of intersection)
- $n$ is also selected by $q_{(k)}$ from $n_j$ (compensation in right-hand side of intersection).

Therefore we now have the following:

$$\begin{aligned}
\mathsf{Pr}(e_i \cap e_j) =& \mathsf{Pr}(n_i \in v'(\mathcal{P})) \times \mathsf{Pr}(n \in \alpha(\mathcal{P}^{n_i}) \mid n_i \in v'(\mathcal{P}))\\
=& \mathsf{Pr}(n_i \in v'(\mathcal{P})) \times \mathsf{Pr}(n \in \alpha(\mathcal{P}_v^{n_i}) \mid n_i \in v'(\mathcal{P}))\\
=& \mathsf{Pr}(n_i \in v'(\mathcal{P})) \times \mathsf{Pr}(n \in \alpha(\mathcal{P}_v^{n_i}))\\
=& [\mathsf{Pr}(n_i \in v(\mathcal{P})) \div \mathsf{Pr}(n_i \in l_m[Q_m](\mathcal{P}_v^{n_i}))] \times\\
& \times\ \mathsf{Pr}(n \in \alpha(\mathcal{P}_v^{n_i})).
\end{aligned}$$

We can take the steps in $\alpha$ because different images in $\mathcal{P}_v$ of the part $l_1/\dots/l_m$ of $v$ are necessarily disjoint. The second reformulation is an immediate consequence of $v' \perp q''$. The third one follows also from it, since $v_{(k)} = l_m[Q_m]$. So $\mathsf{Pr}(e_i \cap e_j)$ can be computed using $v$'s results, with the special representation of Id values allowing us here a simpler formulation for it.

Any conjunction of up to $a$ events can be computed similarly.

**Case $\mathbf{u \geq 1}$.** Let $(l_1, \dots, l_u)$ be the maximal prefix-suffix of the sequence $(l_1, \dots, l_m)$, and assume that there are no predicates on the first $u-1$ nodes of $v$'s last token $t$ of the form: $t =$

$$l_1/l_2/\dots/l_{u-1}/l_u[Q_u]/\dots/l_1[Q_{m-u+1}]/\dots/l_{u-1}[Q_{m-1}]/l_u[Q_m]$$

We describe below the formula for $\mathsf{Pr}(e_i \cap e_j)$, in the case of two events $e_i$ and $e_j$. For $n_i$ ancestor of $n_j$, let $s(i,j)$ denote the number of data nodes from $n_i$ to $n_j$ in $\widehat{\mathcal{P}}$, including these two nodes (we can always get the $s(i,j)$ values from $\widehat{\mathcal{P}}_v^{n_i}$). The formula for the probability of joint events will change slightly (via the $\alpha$ pattern).

If $s(i,j) > m$, the $\alpha$ pattern and the probability formula remain the same as in the case of $u = 0$.

Otherwise, if $s(i,j) \leq m$, let $\alpha$ be defined by the TP$^\cap$-pattern:

$$\alpha = q_{(k)} \cap \texttt{comp}(l_{m-s(i,j)+1}[Q_{m-s(i,j)+1}]/../l_m[Q_m][Id(n_j)], q_{(k)})$$

Then, we can formulate $\mathsf{Pr}(e_i \cap e_j)$ as before:

$$\begin{aligned}
\mathsf{Pr}(e_i \cap e_j) =& \mathsf{Pr}(n_i \in v'(\mathcal{P})) \times \mathsf{Pr}(n \in \alpha(\mathcal{P}^{n_i}) \mid n_i \in v'(\mathcal{P}))\\
=& \mathsf{Pr}(n_i \in v'(\mathcal{P})) \times \mathsf{Pr}(n \in \alpha(\mathcal{P}_v^{n_i}) \mid n_i \in v'(\mathcal{P}))\\
=& \mathsf{Pr}(n_i \in v'(\mathcal{P})) \times \mathsf{Pr}(n \in \alpha(\mathcal{P}_v^{n_i})) = \mathsf{Pr}(n \in \alpha(\mathcal{P}_v^{n_i}))\\
& \times [\mathsf{Pr}(n_i \in v(\mathcal{P})) \div \mathsf{Pr}(n_i \in l_m[Q_m](\mathcal{P}_v^{n_i}))]
\end{aligned}$$

A similar approach can be used to compute any conjunction of up to $a$ events under the assumption that there are no predicates on the first $u-1$ nodes of $v$'s last token. $\square$

We summarize the results of this section with algorithm TPrewrite (Figure 6), which takes as input a TP query $q$ and a set of views $V$ and returns all possible deterministic TP-rewritings $q_r$ which can be complemented by a $f_r$ function, for a probabilistic TP-rewriting.

PROPOSITION 4. TPrewrite *is sound and complete for deciding whether a probabilistic* TP*-rewriting of a query $q$ over views $V$ exists. It runs in* PTime *in the size of the query and views.*

*Remark.* Note that, in the case of probabilistic TP-rewritings, there is a complexity separation between the decision problem for the existence of a rewriting – which is tractable – and the one of executing the alternative access plans based on views – which can be done in EXPTime. Exponential time in the size of the query and views is unavoidable in practice, since TP-query evaluation over

**INPUT** : TP query $q$ and views $V$
**OUTPUT**: Set of TP-rewritings $R$

$R := \emptyset$;
**for** *each* $v \in V$ **do**
  $k := |\texttt{mb}(v)|$,
  $t :=$ last token of $v$,
  $u :=$ size of maximal prefix-suffix in $t$
  **if** $\texttt{comp}(doc(v)/\mathsf{lbl}(v), q_{(k)}) \equiv q$ **then**
    $q^{(k)} :=$ the prefix of $q$ of size $k$
    $v' := v$ w/o predicates at node of depth $k$ ($\texttt{out}(v)$)
    $q'' := \texttt{comp}(\texttt{mb}(q^{(k)}), q^{(k)}_{(k)})$
    **if** $v' \not\perp q''$ **then** continue;
    **if** $\texttt{comp}(doc(v)/\mathsf{lbl}(v), q_{(k)})$ *is restricted* **then**
      $R := R \cup \{\texttt{comp}(doc(v)/\mathsf{lbl}(v), q_{(k)})\}$
    **else if** *no predicates on the first* $u-1$ *nodes in* $t$ **then**
      $R := R \cup \{\texttt{comp}(doc(v)/\mathsf{lbl}(v), q_{(k)})\}$

**Figure 6: Algorithm TPrewrite for probabilistic TP-rewritings**

p-documents (and view extensions) is intractable in query size. We strongly conjecture that the same complexity bounds should remain valid for the evaluation of intersections of tree pattern queries, as in the deterministic case. Although our general formula from Eq. (1), for the probability-retrieving function, can be exponentially large in the size of the view result (by the inclusion-exclusion formulation), it can be reformulated into one that remains tractable in the size of the data, in a rather technical but not very complex manner. For space reasons, these details are omitted.

# 5. TP$^\cap$-REWRITINGS

We consider in this section the problem of view-based rewriting over probabilistic data in the presence of persistent node Ids, using TP$^\cap$-rewritings, i.e., intersections of possibly compensated views. The pattern $q_r$ of a TP$^\cap$-rewriting will now be of the form $\bigcap_{i,j} u_{ij}$, where each $u_{ij}$ is a TP-rewriting over some view $v_j$, i.e., a possibly compensated view.

Let $V = \{v_1, \dots, v_m\}$, with $m = |V|$, be the set of TP views to be used in a rewriting (each $v_i$ contains $q$ or a prefix thereof). Given an candidate plan $q_r = \bigcap_{i,j} u_{ij}$ in TP$^\cap$, verifying that it is a deterministic rewriting of $q$ in TP can be done by verifying $unfold_V(q_r) \equiv q$ (see [10]), which in turn amounts to testing that

*(i)* each TP query in $unfold_V(u_{ij})$ contains $q$, and
*(ii)* $q$ contains the TP$^\cap$ query $unfold_V(q_r)$.

Before discussing TP$^\cap$-rewritings, we recall how one can decide containment and equivalence between a TP query and a TP$^\cap$ one.

## 5.1 Equivalence and Containment for TP$^\cap$

Since $Q$ in TP$^\cap$ is a rewriting for $q$ in TP iff $unfold_V(Q) \equiv q$, deciding whether a TP query $q$ is equivalent to a TP$^\cap$ query $Q$ is a crucial step for our problem. It is known [10] that one can rely on mappings to decide whether $q \equiv Q$. For that, $Q$ can be first equivalently reformulated into the union of TP queries $\cup_i Q_i$, called its possible *interleavings*, which can be exponentially large in $|Q|$. Interleavings capture all the possible ways to *order* or *coalesce* the main branch nodes of queries participating in the intersection.[4] Testing $q \equiv Q$ was shown to be coNP-hard and boils down to testing $q \equiv \cup_i Q_i$, which in turn boils down to testing that (i) for *some* $j$, $q \sqsubseteq Q_j$, and (ii) for *all* $i$, $Q_i \sqsubseteq q$. (This is reminiscent of results from relational databases, on comparing conjunctive queries with unions of conjunctive queries.) We can immediately conclude that the following also holds:

[4] I.e., ways that are not leading to unsatisfiability.

COROLLARY 2. *Deciding the existence of a probabilistic* $\mathsf{TP}^{\cap}$*-rewriting for a* TP *query* $q$ *and* TP *views* $V$ *is* coNP*-hard.*

The equivalence problem was however shown in [10] to be solvable in PTIME when $q$ belongs to a restricted fragment of TP, called *extended skeletons*.

*Extended skeletons.* Informally, this fragment limits the use of $//$-edges in predicates, in the following manner: a token $t$ of a TP query $v$ will not have predicates that have $//$-edges and that may become redundant because of descendants of $t$ and their respective predicates in some interleaving $v$ might be involved in (by intersecting $v$ with some other query). To define extended skeletons we use the following additional terminology: by a *//-subpredicate* $st$ we denote a predicate subtree whose root is connected by a $//$-edge to a linear /-path $l$ that comes from the main branch node $n$ to which $st$ is associated (as in $n[...[.//st]]$). $l$ is called the incoming /-path of $st$ and can be empty. Extended skeletons are patterns $v$ having the following property: for any main branch node $n$ and $//$-subpredicate $st$ of $n$, there is no mapping (in either direction) between the incoming /-path of $st$ and the /-path following $n$ in the main branch (where the empty path is assumed to map in any other path). For example, the expressions $a[b//c//d]/e//d$ or $a[b//c]/d//e$ are extended skeletons, while $a[b//c]/b//d$, $a[b//c]//d$, $a[.//b]/c//d$ or $a[.//b]//c$ are not.

This TP sub-fragment does not restrict in any way the use of $//$-edges in the main branch or the use of predicates with /-edges only.

As our focus in this paper is on efficient algorithms for view-based rewriting, it is thus natural to ask if over probabilistic data this problem remains tractable, when we deal with extended skeletons under persistent node Ids. *Our general approach hereafter will be to describe decision and evaluation techniques that are sound and complete when applied to any queries and views, but may depend, unavoidably, on equivalence tests involving* $\mathsf{TP}^{\cap}$*-queries. Therefore, their complexity will depend on the one of such tests.*

## 5.2 Using Pairwise c-Independent Views

As for TP-rewritings, we present our results starting with the assumption that a deterministic rewriting $q_r$ has been found. Without loss of generality, let us first assume that $q_r$ consists only of intersected views (plans with possibly *compensated* views are discussed in Section 5.4). W.l.o.g., $q_r$ can be of the form $q_r = doc(v_1)/v_1 \cap \cdots \cap doc(v_m)/v_m$ and, necessarily, $q \sqsubseteq v_i$ for all $v_i$.

We first give some intuition on the possible construction of the probability component of the rewriting, $f_r$: for a given node $n \in q(\mathcal{P})$, since each view $v_i$ gives a probability value, $\mathsf{Pr}(n \in v_i(\mathcal{P}))$, and since we are interested in the probability of the intersection thereof, we might be tempted to try what is arguably the most intuitive definition for $f_r$ here, the one which would simply combine by *multiplication* the values $\mathsf{Pr}(n \in v_i(\mathcal{P}))$. There are however two issues with this straightforward $f_r$ candidate.

The first issue is probabilistic dependencies. We have introduced the notion of c-independence in the previous section, which can guarantee that the existence of some embedding of a view $v_i$ in a given document does not depend – w.r.t predicate conditions – on the existence (or non-existence) of some embedding of another view $v_j$ in this document. We will see now that, for pairwise c-independent views, a function $f_r$ based on multiplication of these views' probabilities can be built.

The second issue has to do with an adjustment for a probability term that appears in each of the views' probability values. More precisely, for each node $n$ that appears in $q(\widehat{\mathcal{P}})$ and, consequently, appears in each $v_1(\widehat{\mathcal{P}}), \ldots, v_m(\widehat{\mathcal{P}})$, we have $m$ probability values $\mathsf{Pr}(n \in v_i(\mathcal{P}))$. Furthermore, each value $\mathsf{Pr}(n \in v_i(\mathcal{P}))$ can be seen as the product of two distinct probability terms:

*(i)* the (appearance) probability of $n$ being part of a possible world of $\widehat{\mathcal{P}}$, denoted $\mathsf{Pr}(n \in \mathcal{P})$,

*(ii)* the probability of $n$ being *selected* by $v_i$ in a possible world *in which* $n$ *is known to appear*, denoted in the following $\mathsf{Pr}(n \in v_i(\mathcal{P}) \mid n \in \mathcal{P})$.

Note that the first term is independent of any particular view to whose result we may be referring, as it only depends on the document itself (this is reflected by our notation). We can thus write for each $v_i$ and node $n$ that

$$\mathsf{Pr}(n \in v_i(\mathcal{P})) = \mathsf{Pr}(n \in \mathcal{P}) \times \mathsf{Pr}(n \in v_i(\mathcal{P}) \mid n \in \mathcal{P}).$$

Given a deterministic rewriting $q_r$ of $q$ formed by pairwise c-independent views $v_1, \ldots, v_m$, for a node $n \in q(\mathcal{P})$), we would thus have as the overall product the following formulation:

$$\prod_i \mathsf{Pr}(n \in v_i(\mathcal{P})) = \mathsf{Pr}(n \in \mathcal{P})^m \times \prod_i \mathsf{Pr}(n \in v_i(\mathcal{P}) \mid n \in \mathcal{P}). \quad (3)$$

Therefore, in Eq. (3) we account for the probability $\mathsf{Pr}(n \in \mathcal{P})$ too many times, once for each view that participates in the rewriting, although we should account for it exactly once. By dividing Eq. (3) with $\mathsf{Pr}(n \in \mathcal{P})^{m-1}$, we can correct this overuse of $n$'s appearance probability, obtaining the following $f_r$ formula when the views $v_1, \ldots, v_m$ are pairwise c-independent:

$$f_r(n) = \mathsf{Pr}(n \in \mathcal{P}) \times \prod_i \mathsf{Pr}(n \in v_i(\mathcal{P}) \mid n \in \mathcal{P}) \quad (4)$$

$$= \prod_i \mathsf{Pr}(n \in v_i(\mathcal{P})) \div \mathsf{Pr}(n \in \mathcal{P})^{m-1}. \quad (5)$$

Each c-independent view $v_i$ gives us $\mathsf{Pr}(n \in v_i(\mathcal{P}) \mid n \in \mathcal{P})$, but there is still one missing ingredient in order to be able to compute $f_r$ as in Eq. (4): $n$'s appearance probability value, $\mathsf{Pr}(n \in \mathcal{P})$.

We can prove the following:

LEMMA 3. $\mathsf{Pr}(n \in \mathcal{P})$ *is computable from* $\widehat{\mathcal{P}}_{v_1}, \ldots, \widehat{\mathcal{P}}_{v_m}$ *iff there exists one* $v_i$ *verifying* $\mathtt{mb}(q) \sqsubseteq v_i$.

Using Lemma 3, we sum up the positive results of this section in the next theorem:

THEOREM 3. *Let* $q$ *be a* TP *query,* $v_1, \ldots, v_m$ *a set of pairwise c-independent* TP *views s.t. there exists a* $v_i$ *satisfying* $\mathtt{mb}(q) \sqsubseteq v_i$*. Let* $q_r$ *be a deterministic* $\mathsf{TP}^{\cap}$*-rewriting of* $q$*, of the form*

$$q_r = doc(v_1)/v_1 \cap \cdots \cap doc(v_m)/v_m.$$

*Then,* $(q_r, f_r)$ *with* $f_r$ *as in Eq. (4), is a probabilistic* $\mathsf{TP}^{\cap}$*-rewriting of* $q$ *over* $V$.

PROOF SKETCH. We refer the reader to [11], for the formal proof based on the material that precedes in this section. □

EXAMPLE 15. *Consider the following (compensated) view for* $v^2_{\mathrm{BON}}$*:* $v = \mathtt{comp}(doc(v^2_{\mathrm{BON}})/bonus, q_{(3)})$*. Clearly the views* $v^1_{\mathrm{BON}}$ *and* $v$ *are c-independent, and* $q_{\mathrm{RBON}} = v^1_{\mathrm{BON}} \cap v$*. According to Theorem 3, the probability* $\mathsf{Pr}(n_5 \in q_{\mathrm{RBON}}(\mathcal{P}_{\mathrm{PER}}))$ *equals* $0.75 \times 0.9 \div 1$ *by the expression*

$$\mathsf{Pr}(n_5 \in v^1_{\mathrm{BON}}(\mathcal{P}_{\mathrm{PER}})) \times \mathsf{Pr}(n_5 \in v(\mathcal{P}_{\mathrm{PER}})) \div \mathsf{Pr}(n_5 \in \mathcal{P}_{\mathrm{PER}}).$$

*For other* $n_i$, $\mathsf{Pr}(n_i \in q_{\mathrm{RBON}}(\mathcal{P}_{\mathrm{PER}}))$=$\mathsf{Pr}(n_i \in v^1_{\mathrm{BON}}(\mathcal{P}_{\mathrm{PER}}))$=0.

We next show that, even for very limited ($//$-free) input queries and views $V$, it is hard to decide the existence of a subset of pairwise c-independent views from $V$ on which a rewriting as in Theorem 3 can be built. This implies that, for extended skeletons as well, it is hard to find $\mathsf{TP}^{\cap}$-rewritings by Theorem 3's approach.

THEOREM 4. *Let* TP *query* $q$ *and* TP *views* $V$ *be without* $//$*-edges. Then, deciding whether there exists a* $\mathsf{TP}^{\cap}$*-rewriting of* $q$ *using only pairwise c-independent views from* $V$ *is* NP*-hard.*

PROOF SKETCH. We prove this by reduction from the problem of $k$-DIMENSIONAL PERFECT MATCHING: *Given a k-hypergraph $H = (U; E)$ with $s = |U|$ and $m = |E|$, is there a subset $S \subseteq E$ of $s/k$ hyperedges such that each vertex of $U$ is contained in exactly one hyperedge of $S$?*

Let $U = \{u_1, \ldots, u_s\}$ and let $E = \{e_1, \ldots, e_m\}$ denote the nodes and edges of the k-dimensional hypergraph $H$.

We build an input query $q$ of the form $q = a[1]/a[2]/\ldots/a[s]//b$. For each edge $e_i \in E$, we build a view $v_i$ as follows: a sequence of $s$ $a$-labeled nodes separated by $/$-edges, followed by a $//$-edge and then a $b$-labeled node. On the $a$-nodes, the predicates corresponding to the vertices of $e_i$ are present. For example, for an edge $e_i = (1, 2, 3)$ (for $k = 3$) we would construct the following view:

$$v_i = a[1]/a[2]/a[3]/a\ldots/a//b.$$

Now, if a perfect matching for $H$ exists, one can notice that the views corresponding to the edges of this matching should be c-independent. Their intersection gives an equivalent rewriting both in the absence and in the presence of probabilities. Vice-versa, if we find a rewriting of $q$ using views, this means that we have a set of views which are c-independent and cover all the predicates of $q$. But this amounts to finding a perfect matching for $H$. □

## 5.3 Using View Decompositions

Theorem 4 shows that it may not be possible to find a $\text{TP}^\cap$-rewriting based on c-independent views efficiently. It may however be possible to build probabilistic rewritings, without requiring pairwise c-independence. We describe in this section a general, sound and complete approach, which attempts to build a *system of probability values* from the views' results, even in the presence of dependent views. We first give the intuition behind it with an example.

EXAMPLE 16. *Consider the input query $q = a[1]/b[2]/c[3]/d$, and the views $v_1 = a[1]/b/c[3]/d$, $v_2 = a/b[2]/c[3]/d$, $v_3 = a[1]/b[2]/c/d$, and $v_4 = a//d$.*

*Note that the first three views are not c-independent with each other. Their intersection does yield a deterministic $\text{TP}^\cap$-rewriting of q (in fact, $v_1$ and $v_2$ would suffice for a deterministic rewriting of q). Moreover, $v_4$, which gives us the probability $\mathsf{Pr}(n \in \mathcal{P})$ for any node n in q's result, would be considered redundant in the deterministic setting (it contains the other three views).*

*For a probabilistic rewriting, it remains to specify the probability function $f_r$. However, due to probabilistic dependencies between the views, it is no longer possible to simply multiply the individual probabilities from $v_1, v_2, v_3$ ($v_4$ gives the values $\mathsf{Pr}(n \in \mathcal{P})$).*

*However, one could retrieve the values $\mathsf{Pr}(n \in q(\mathcal{P}))$, by slightly more involved arithmetic manipulations, starting from the observation that these can be seen as the product of four independent probability values: $\mathsf{Pr}(n \in \mathcal{P})$ and one value for each of the three predicates. With a slight abuse of notation, for $j \in \{1, 2, 3\}$ let $\mathsf{Pr}(j)$ denote the probability of the predicate $[j]$ to match at the corresponding depth. The following system of equations can then be built straightforwardly, for each n such that $\mathsf{Pr}(n \in q(\mathcal{P})) > 0$ :*

$$\mathsf{Pr}(n \in v_1(\mathcal{P})) = \mathsf{Pr}(n \in \mathcal{P}) \times \mathsf{Pr}(1) \times \mathsf{Pr}(3),$$
$$\mathsf{Pr}(n \in v_2(\mathcal{P})) = \mathsf{Pr}(n \in \mathcal{P}) \times \mathsf{Pr}(2) \times \mathsf{Pr}(3),$$
$$\mathsf{Pr}(n \in v_3(\mathcal{P})) = \mathsf{Pr}(n \in \mathcal{P}) \times \mathsf{Pr}(1) \times \mathsf{Pr}(2),$$
$$\mathsf{Pr}(n \in v_4(\mathcal{P})) = \mathsf{Pr}(n \in \mathcal{P}),$$

*which would allow us to obtain easily the probability values*

$$\mathsf{Pr}(n \in q(\mathcal{P})) = \mathsf{Pr}(n \in \mathcal{P}) \times \mathsf{Pr}(1) \times \mathsf{Pr}(2) \times \mathsf{Pr}(3),$$

*provided the system allows an unique solution for the unknown value of $\mathsf{Pr}(n \in q(\mathcal{P}))$. This condition can be verified independently of any node n (i.e, it is not data dependent).*

*Outline of the general approach.* The idea illustrated by Example 16 can be generalized into an algorithm that applies to any query and views, without being data dependent. In principle, we will still rely on probability terms that are independent, but at a more fine-grained level. More precisely, we will decompose the set of views $V = \{v_1, \ldots v_m\}$ into a set of pairwise c-independent *view decompositions* (in short d-views; these are queries from TP as well) denoted $w_1, \ldots, w_s$, and then use d-views instead of the given ones. The major difference w.r.t. the setting of Theorem 3 is that now we will not have the explicit probabilities of d-views' results, but only some combinations thereof (from the results of the given views), in a non-homogenous linear system. Based on this system, we will then describe a decision procedure (sound and complete) for the existence of the $f_r$ function, running in $\mathsf{PTime}$ in the size of the query and of the initial set of views.

We start by describing how we move from the initial set of views to the set of d-views $w_1, \ldots, w_s$, by the following four steps.

We can see each view $v_i$ as being of the form $v_i = ft_i//m_i//lt_i$, where $ft_i$ is the first token, $lt_i$ is the last token, and $m_i$ denotes the rest ($m_i$ may be empty; if $m_i$ is empty, $lt_i$ may also be empty).

*Step* 1. For each $v_i$, build the TP queries $w_i^1, \ldots, w_i^{s_i}$ as follows:

*(i)* $|\mathtt{mb}(ft_i)|+|\mathtt{mb}(lt_i)|$ queries: one query for each main branch node $n$ of either $ft_i$ or $lt_i$, obtained from $v$ by removing all predicates from $v$ except the ones on $n$.

*(ii)* One query of the form $\mathtt{mb}(ft_i)//m_i//\mathtt{mb}(lt_i)$, i.e., obtained from $v_i$ by keeping only the predicates of the $m_i$ part.

*Step* 2. For each view $v_i$ and its $w_i^j$ queries obtained at the previous step, repeat until no change occurs the following: replace any two queries $w_i^x, w_i^y$ s.t. $w_i^x \not\perp w_i^y$ by their intersection $w_i^x \cap w_i^y$.

*Step* 3. Replace each query obtained at the previous step by its intersection with the linear query $\mathtt{mb}(q)$.

*Step* 4. Across the $m$ sets of queries obtained at the previous step, group the queries into equivalence classes (by query equivalence). Then, by introducing one d-view name for each equivalence class, output the final set of d-views $\{w_1, \ldots, w_s\}$.

For each of the initial views $v_i$, let $W_i \subseteq \{w_1, \ldots w_s\}$ denote the d-views into which it is decomposed. We can now write the following equation for each view $v_i$:

$$\mathsf{Pr}(n \in v_i(\mathcal{P})) = \mathsf{Pr}(n \in \mathcal{P}) \times \prod_{w_j \in W_i} \mathsf{Pr}(n \in w_j(\mathcal{P}) \mid n \in \mathcal{P}) \quad (6)$$

For the input query $q$, let $W_q \subseteq \{w_1, \ldots w_s\}$ denote the d-views into which $q$ can be decomposed. We have an additional equation:

$$\mathsf{Pr}(n \in q(\mathcal{P})) = \mathsf{Pr}(n \in \mathcal{P}) \times \prod_{w_j \in W_q} \mathsf{Pr}(n \in w_j(\mathcal{P}) \mid n \in \mathcal{P}). \quad (7)$$

Let $\mathcal{S}(q, V)$ denote the non-homogenous system of $m + 1$ linear equations that can be obtained from Eq. (6) and Eq. (7), by taking the logarithm. Note that $\mathcal{S}(q, V)$ has $s + 2$ variables: $s$ variables corresponding to $\mathsf{Pr}(n \in w_j(\mathcal{P}) \mid n \in \mathcal{P})$ terms, one variable for $\mathsf{Pr}(n \in \mathcal{P})$, and one *distinguished* variable for $\mathsf{Pr}(n \in q(\mathcal{P}))$.

We are now ready to present our main results for probabilistic $\text{TP}^\cap$-rewritings :

THEOREM 5. *Let q be a* TP *query, $V = \{v_1, \ldots, v_m\}$ be a set of* TP *views containing q. Let $q_r$ be a deterministic* $\text{TP}^\cap$*-rewriting of q, of the form $q_r = doc(v_1)/v_1 \cap \cdots \cap doc(v_m)/v_m$. There exists a probabilistic* $\text{TP}^\cap$*-rewriting, of the form $(q_r, f_r)$, if $\mathcal{S}(q, V)$ admits an unique solution for $\mathsf{Pr}(n \in q(\mathcal{P}))$. Unless $\mathtt{mb}(q)$ has only $/$-edges, such a probabilistic* $\text{TP}^\cap$*-rewriting exists only if $\mathcal{S}(q, V)$ admits an unique solution for $\mathsf{Pr}(n \in q(\mathcal{P}))$.*

PROOF SKETCH. For each of the views participating in $q_r$, the decomposition into a product of independent terms by Eq. (6) is sound. Moreover, it is the maximal one (w.r.t. number of terms) or, put otherwise, the most fine-grained, since dependencies between tests of the views must occur within the same d-view. Any other reformulation of the view probabilities is necessarily either subsumed by $\mathcal{S}(q, V)$ or equivalent to it modulo renaming of variables.

In particular, predicates of the same node must be part of the same d-view (the probabilities for them to match are obviously not independent). Note that we can refer to predicates of the first or last tokens of views - and their probability to match - *unambiguously*, since the main branch nodes of these tokens are unambiguously identified on the path from the root of the p-document to some result node $n$ selected by $q$. But this is not the case for predicates of the $m_i$ parts of views, and the reason we need to consider them "in bulk", by a single $w_i^j$ expression corresponding to all of them.

At Step 3 (intersection with $\mathtt{mb}(q)$), we simply explicit the fact that nodes $n$ we are interested in must be found on the path matching $\mathtt{mb}(q)$, as they verify $\Pr(n \in q(\mathcal{P})) > 0$. Omitting this step would keep the approach sound, but may cause us to miss opportunities to reuse the same variable across distinct views (and ultimately find the $f_r$ function). (For a detailed proof, see [11].) □

PROPOSITION 5. *Let $q$ be a* TP *query and $V = \{v_1, \ldots, v_m\}$ be a set of* TP *views s.t. there exists a deterministic* TP$^\cap$*-rewriting of $q$, of the form $q_r = doc(v_1)/v_1 \cap \cdots \cap doc(v_m)/v_m$. Testing whether the system $\mathcal{S}(q, V)$ admits an unique solution for $\Pr(n \in q(\mathcal{P}))$ can be done in* PTime*, modulo* TP$^\cap$*-equivalence tests.*

PROOF SKETCH. Finding a deterministic rewriting requires an equivalence test of the kind described in Section 5.1, hence a worst-case exponential step.

Regarding the $f_r$ component for a probabilistic rewriting, in the $\mathcal{S}(q, V)$ construction, by the way intersections of patterns are constructed at Step 2, these TP$^\cap$ queries reduce trivially to equivalent TP ones (they are *union-free*, in the terminology of [8]). So we can safely assume that each run of Step 2 deals only with tree patterns, instead of intersections thereof. At Step 4, the d-views are obtained based on equivalence tests, which may involve TP$^\cap$ queries.

Then, testing if $\mathcal{S}(q, V)$ admits an unique solution for $\Pr(n \in q(\mathcal{P}))$ can be done in polynomial time by straightforward linear algebra manipulations. Note that this does not necessarily mean that $\mathcal{S}(q, V)$ admits an unique solution for all its variables. □

## 5.4 Dealing with Compensated Views

We consider in this section general TP$^\cap$-rewritings that, before performing the intersection step, might compensate (some of) the views. We show that rewriting in this new setting can be reduced to the one discussed in the previous section, by relying also on the results of Section 4. This allows us to reuse the same PTime algorithm and to find strictly more rewritings, namely those that would not be feasible without compensation.

The general approach will be the following. Starting from the given set of views $V$, all containing the input query $q$ or a prefix thereof, we will expand $V$ into $V'$ by adding to it all possible compensated views of the form $\mathtt{comp}(v, q_{(a)})$, for $v \in V$ and $a$ being a depth in the range 1 to $|\mathtt{mb}(q)|$. As in [8], the views of $V'$ will then be used to build what we call *the canonical deterministic plan* $q_r = \bigcap_{v_i \in V'} doc(v_i)/v_i$. For a probabilistic rewriting, among the views of $V'$, we select the subset $V''$ of those that (i) either were originally in $V$, or (ii) verify certain conditions that ensure their result probabilities can indeed be computed from the initial results of the view they were constructed on (using the decision procedure described in Section 4).

Algorithm TPIrewrite (see Figure 7) details this approach, and it represents a decision procedure for finding TP$^\cap$-rewritings based on possibly compensated views. It takes as input a TP query $q$ and

**INPUT** : TP query $q$ and views $V$
**OUTPUT**: canonical rewriting $q_r$

$R := \emptyset,\ V' := V,\ V'' := V$

*Prefs* $:= \{(v_i, a) \mid q^{(a)} \sqsubseteq v_i$, *for* $q^{(a)}$ *being the prefix of size* $a$ *of* $q\}$;

**for** *each* $v \in V$ **do**
  $k := |\mathtt{mb}(v)|$,
  $t :=$ last token of $v$,
  $u :=$ size of maximal prefix-suffix in $t$
  **for** *each* $(v, a) \in$ *Prefs* **do**
    $V' := V' \cup \{\mathtt{comp}(v, q_{(a)})\}$
    $v' := v$ w/o predicates at node of depth $k$ ($\mathtt{out}(v)$)
    $q'' := \mathtt{comp}(\mathtt{mb}(v), q_{(a)}^{(a)})$
    **if** $v' \not\sqsubseteq q''$ **then** continue;
    **if** $\mathtt{comp}(doc(v)/\mathsf{lbl}(v), q_{(a)})$ *is restricted* **then**
      $V'' := V'' \cup \{\mathtt{comp}(v, q_{(a)})\}$
    **else if** *no predicates on the first* $u - 1$ *nodes in* $t$ **then**
      $V'' := V'' \cup \{\mathtt{comp}(v, q_{(a)})\}$

$q_r = \bigcap_{v_i \in V'} doc(v_i)/v_i$
**if** *unfold*$(q_r) \equiv q$ **then**
  **if** $\mathcal{S}(q, V'')$ *has unique solution for* $\Pr(n \in q(\mathcal{P}))$ **then**
    return **true**

**Figure 7:** TPIrewrite **for probabilistic** TP$^\cap$**-rewritings**

a set of TP views $V$ and returns the canonical deterministic TP$^\cap$-rewriting $q_r$, whenever the $f_r$ function can also be built.

We can prove the following main result for TP$^\cap$-rewritings:

PROPOSITION 6. TPIrewrite *is sound for testing if a probabilistic* TP$^\cap$*-rewriting of a query $q$ over views $V$ exists. It is also complete, unless $\mathtt{mb}(q)$ has only $/$-edges.*

*Modulo* TP$^\cap$*-equivalence tests,* TPIrewrite *runs in* PTime *in the size of the query and views.*

Potentially expensive equivalence tests may be performed when we verify whether $q_r$ is a deterministic rewriting (step before last in TPIrewrite) or in the construction of the $\mathcal{S}(q, V)$ system. But these can be performed efficiently when we deal with the restricted fragment of extended skeletons.

COROLLARY 3 (OF PROP. 6). *If the views $V$ and query $q$ are extended skeletons,* TPIrewrite *runs in* PTime *in the size of the query and views.*

*Remark.* Note that the evaluation of the rewriting on the view extensions might require first the evaluation of any compensated view used in $q_r$ and $f_r$, hence may require exponential time in the size of the query and views, which is not surprising given that the evaluation of TP queries over probabilistic XML is intractable.

# 6. OTHER RELATED WORK

There is a rich literature on query rewriting using views for deterministic XML data. XPath rewriting using only one view [36, 25, 37] or multiple views [6, 34, 5, 8, 26] was the topic of several studies. They differ in the completeness guarantees they provide, or the assumptions they rely on.

Some join-based rewriting methods either give no completeness guarantees [6, 34] or can do so only if there is extra knowledge about the structure and nesting depth of the XML document [5]. Others can only be used if the node Ids are in a special encoding that accounts for structural information [34]. Rewriting more expressive XML queries using views was studied in [14, 18, 29].

[17] studies query answering using views for relational probabilistic data. There is little work dealing with the optimization of

query answering for probabilistic XML. A system that uses relational probabilistic databases for storying and managing probabilistic XML is proposed in [20]. Approximate computation of probabilities for tree-pattern queries over probabilistic XML is studied in [22, 33]. We presented some preliminary results on rewriting for probabilistic XML at a workshop without formal proceedings [12].

# 7. CONCLUSION AND FUTURE WORK

Our work is the first to address the problem of answering queries using views over probabilistic XML. Since in a probabilistic setting queries return answers with probabilities, view-based rewriting goes beyond the classic problem of retrieving answers from XML views. Thus, the new challenge raised in the probabilistic setting is to find probability-retrieving functions that can access only view results, while being able to compute the probabilities of answers.

We identified large classes of XPath queries – with child and descendant navigation and predicates – for which there are efficient (PTime) algorithms, considering the rewriting problem under the two possible semantics for XML query results: with persistent node identifiers and in their absence. Accordingly, we considered rewritings that can exploit a single view, by means of compensation, and rewritings that can use multiple views, by means of intersection. Recall that (direct) query answering for probabilistic XML model considered here is also polynomial in data and intractable in query complexity [22].

All our results are practically interesting, as they allow expressive queries and views, with descendant navigation and path filter predicates, and our decision procedures are based on easily verifiable criteria on the query and views. For both semantics, the evaluation of an alternative plan is no more expensive then query evaluation over probabilistic XML. Moreover, rewritings based solely on intersection would require only the computation of the $f_r$ function, either by a product formula or by solving a $\mathcal{S}(q, V)$ system, operations that should cost significantly less than the dynamic programming approach for query evaluation over the original data [22]. Even for plans that do use compensation (which may require TP-evaluation over view extensions), the costs should be reduced in practice, especially if extensions are much smaller that the original p-document. We intend to evaluate the impact of these techniques in practice, in a probabilistic XML management system. Also, heuristics for choosing the views that participate in a rewriting, tailored to the setting of probabilistic XML data, may represent valuable optimizations. Beyond the natural choice of just caching the probabilistic results, keeping and exploiting for rewritings a sort of *why-provenance* of probability values is also an interesting direction for future research.

Another possible direction for future work is to broaden the setting to other models for probabilistic XML data. The p-documents studied in this paper have local probabilistic dependences, while there are models allowing for more complex probabilistic interactions between remote fragments of data [32]. For these types of data, query answering is intractable (already in data complexity) and it would be interesting to see under which conditions we can gain tractability by relying on views.

*Acknowledgements.* Cautis was supported by the French ANR project DataRing and EU project ARCOMEM (FP7-ICT-270239). Kharlamov was supported by ERC FP7 grant Webdam (n. 226513), EU project ACSI (FP7-ICT-257593), EPSRC grant EP/G004021/1.